**Barkhausen noise in the columnar hexagonal organic ferroelectric BTA**

*Andrey Alekseevich Butkevich, Fabian T. Thome, Toni Seiler, Marcel Hecker, Martijn Kemerink*

Andrey Alekseevich Butkevich, Fabian T. Thome, Toni Seiler, Marcel Hecker, Martijn Kemerink

Institute for Molecular Systems Engineering and Advanced Materials, Im Neuenheimer Feld 225, Heidelberg, 69120, Germany.

Email: martijn.kemerink@uni-heidelberg.de

Keywords: BTAs, crackling noise, kinetic Monte Carlo simulations, ferroelectric switching

Upon a polarization reversal within a ferroelectric material, one stable state changes into another which is typically described by a progression of switching events of smaller fractions of the material. These events give rise to crackling or Barkhausen noise and follow a characteristic distribution in their sizes. Barkhausen noise has been studied to better understand the switching processes of ferroelectrics and has been applied for inorganic ferroelectric materials and perovskites. In this work, we present results from kinetic Monte Carlo simulations investigating the switching process of the small organic molecular ferroelectric benzene-1,3,5-tricarboxamides (BTAs). For temperatures below 175 K and sufficiently strong structural disorder, the system exhibits self-organized critical behavior; for higher temperatures, a creep regime is entered. Our extracted power-law exponents are smaller than those typically measured in inorganic crystals and ceramics which indicates that in the more disordered material BTA larger spanning avalanches are possible. The system was experimentally investigated with a high-sensitivity setup. No Barkhausen noise was observed which is consistent with the simulated event sizes, lying several orders beneath the noise threshold of the experimental setup. This finding corroborates the notion that switching in BTA progresses along the 1D columns in the hexagonal liquid crystal lattice, with little coupling between the columns that could give rise to larger lateral avalanches.

## 1. Introduction

While crackling – or Barkhausen – noise appearing during ferroelectric polarization reversal has been thoroughly investigated, most previous studies were confined to inorganics and specifically perovskites.[1–7] It was established that different inorganic ferroelectric materials indeed show Barkhausen noise in their switching processes. Statistical analyses showed that power-law behavior is present in multiple quantities describing the switching process, although the expected exponential cut-off at large event sizes is not always present. The recovered critical exponents varied between different materials, depending on the strength of correlations present, but the overall similar behavior suggests universality in systems exhibiting Barkhausen noise.[2–4,6,7] The only investigation into thermal Barkhausen noise of organic ferroelectrics was performed on PVDF thin films where the measured current peaks are generated due to the reordering of dipoles while crossing the Curie-temperature and clear deviations from a continuous switching process with different shapes were observed [8]. As such, univocal evidence of critical behavior in the switching of organic ferroelectric materials is still lacking. Observation, or the lack thereof, would, however, provide important information about the microscopic details of switching in organic ferroelectrics, including the size of and the interaction strength between the switching entities.

If a physical system is driven, meaning that it is constantly pushed out of equilibrium by a, typically slowly increasing, external force, one of the possible responses can be crackling noise, which constitutes the transition between the snapping and popping. The former is characterized by a singular large event and the latter is given by many smaller events of similar amplitude. In the intermediate – the crackling – case, the occurrence of sudden and jerky events whose corresponding amplitudes, durations and energies are typically power-law distributed on a large scale arises.[9] Crackling is interpreted as a dynamical critical phenomenon.[9,10] In the case of ferroelectrics, one manifestation of crackling noise is Barkhausen noise, which arises from the irregular movement of domain walls through the sample.[11]

The mean field plasticity model was originally applied to materials under shear stress, displaying step-like stress-strain curves that showed material-independent power-law behavior.[12] Considering the property of universality in critical systems, it was found that the model could also be related to switching process in soft ferroic materials.[13] This process would then be described as a depinning transition where the local failure stress $\tau_{\mathrm{f}}$ has to be overcome for a slip to start that then continues until the local stress is below the sticking stress $\tau_{\mathrm{s}}$. Due to disorder, both quantities vary depending on the location inside the material. The slip itself can cause further slips in the surrounding area, leading to an avalanche propagating through the material as the external force $F$ causes the stress to slowly increase and the medium to deform elastically. Applying a mean field theory, the evolution of the accumulated slip $u(r,t)$ is described as

$$\eta \frac{\partial u(r,t)}{\partial t} = F + \sigma_{\mathrm{int}}(r,t) - f_{\mathrm{R}}\big(u, r, u(r, t' < t)\big), \qquad (1)$$

with the friction coefficient $\eta$, the failure stress distribution $f_R$ and the shear stress $\sigma_{\mathrm{int}}$. The solution of **Equation 1** predicts, in accordance with the domain wall propagation model, a power-law distribution of all sizes with an exponential cut-off.[14] The Mean Field Plasticity model establishes that the critical exponents are stress-dependent. For data collected over large stress ranges, the usual distributions are integrated over the applied force $F$,[15] causing exponents to increase as the biggest avalanches appear at critical stress. The resulting critical exponents of the power-law describing the event size distribution $\tau$ and energy distribution $\epsilon$ are 1.5 and $4/3$, respectively.[12] For ferroic materials, the crackling events are measured over the whole (stress integrated) hysteresis loop and compared to events measured around the coercive field in analogy to stress.[16] The stress integrated mean field values for the critical exponents are given by $\tau = 2$ and $\epsilon = 5/3$.

In this work, the prototypical small organic molecular ferroelectric benzene-1,3,5-tricarboxamide (BTA) is investigated. It is a columnar hexagonal liquid crystal and hence combines the properties like flexibility and easy-processability of a liquid and anisotropy for macroscopic polarization of a crystal. Furthermore, the basic BTA molecule discussed here has many structural derivatives with quantitatively different properties and potential applications.[17] The herein investigated BTA-$C_8$ is representative of the larger family of materials. BTAs consist of a benzene core with three attached amide groups and alkyl tails of varying length, as shown in **Figure 1a**. BTAs tend to self-assemble into columnar stacks via π-π stacking and hydrogen bonding, hence making it a supramolecular columnar discotic which possesses a cooperative effect, meaning that longer stacks generally become more stable and have larger dipole moments per monomer, which is indicated in **Figure 1b**.[18–21] The ferroelectric character of BTA is based on the dipole moments of the amide groups along the columns forming a macrodipole and has been studied and proven by several experimental studies.[17,20,22,23] It has been established that the polarization switching relies on the reorientation of the amide groups and thus on sufficiently flexible side chains.[24] Despite the high grade of organization, BTA systems show a high amount of disorder due to amorphous regions in the material and defects in the molecular columns.

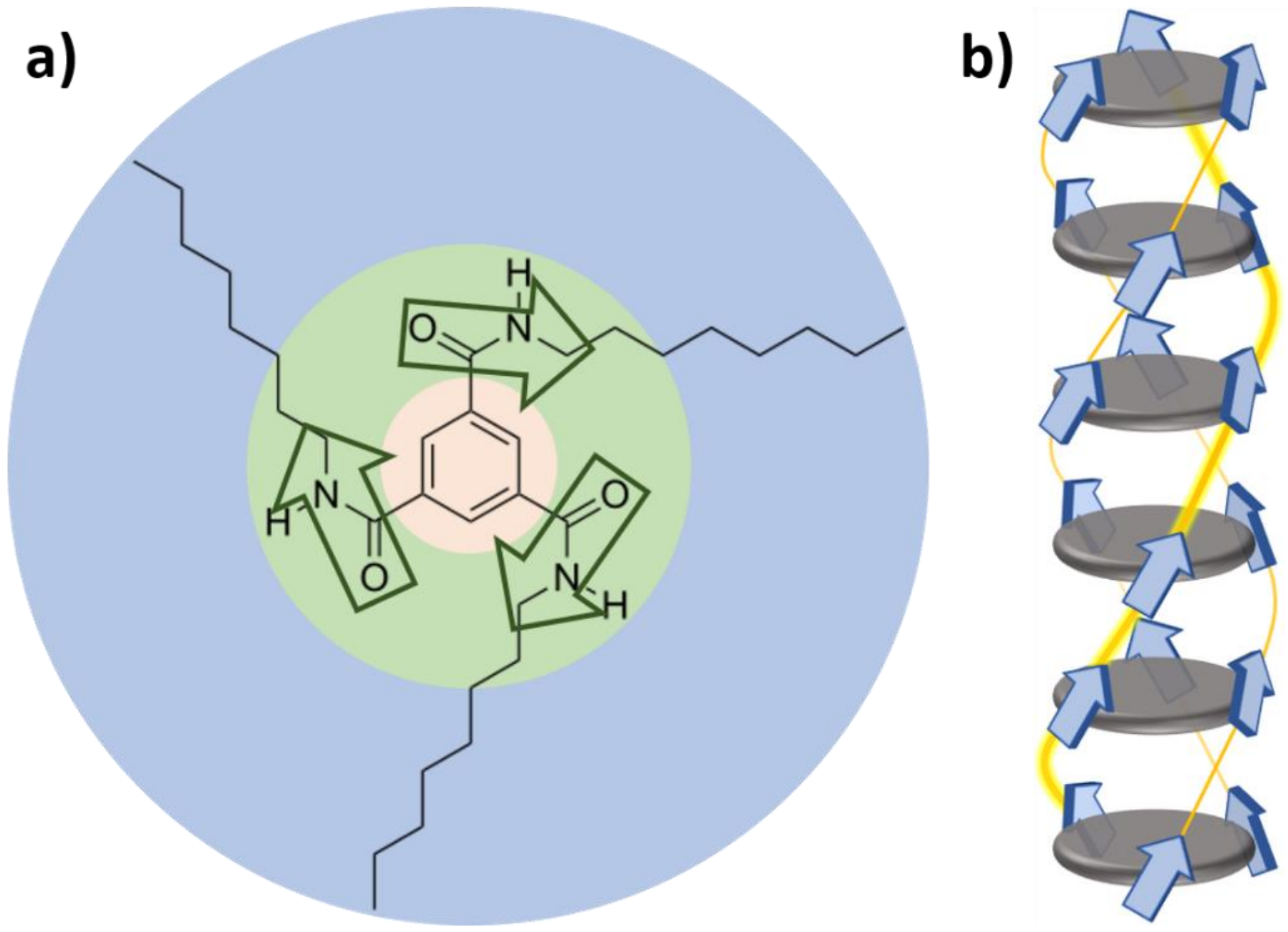

**Figure 1:** The **a)** primary and **b)** tertiary structure of a BTA, in this case with a $C_8$ alkyl tail. The BTA molecule consists of a benzene core (yellow), three amide groups (green), each with a dipole moment (indicated by arrows), and an attached alkyl tail (blue). The dipole moments contribute to a large macro dipole within a BTA column.

Previously, we have hypothesized that the structural disorder in BTA gives rise to small microdomains that strongly interact internally but only weakly with their surroundings.[25] As such, they could be treated as the hysterons in a Preisach model, which allowed to reproduce the shape of the hysteresis curve as well as the dispersion in the switching kinetics. As the Preisach model neglects all interactions between the hysterons beyond the offset on the Preisach plane, the question whether any relevant interaction occurs between the hysterons, that were identified as sub-columns in hexagonal columnar morphology, is still open.

In this work, the fluctuations that naturally emerge in kMC simulations are used to simulate crackling noise. The analysis thereof suggests a universal character of irreversible switching of BTA which occurs via critical dipole avalanches of different sizes as further discussed below. The avalanches propagate along the BTA columns. The power-law exponents extracted from the size and energy distribution functions are seemingly invariant to the disorder of the system and agree well with the mean field theory. While for higher temperatures the exponents are subject to thermal effects, which is attributed to creep behavior, the exponents observed at lower temperatures are robust, in agreement with self-organized criticality. Experimentally, no Barkhausen noise was measured in BTA samples which is attributed to the current signals produced by switched dipoles being too low to be observed with the noise threshold of the used measuring setup.

## 2. Model

In order to study the critical behavior and crackling noise, both processes on the macroscopic scale and molecular level have to be considered. This is achieved via a kinetic Monte Carlo (kMC) simulation that focusses on electrostatic interactions; other, e.g. steric, interactions are accounted for by a prescribed morphology and especially the rotation rates and the angles of the amide dipoles. In previous work we found these parameters to be the most relevant ones and, importantly, sufficient to describe the thermally driven switching in BTA.[25–27] Switching rates, based on classical energy calculations, determine which dipoles are likely to flip and thus the evolution of the system. Additionally, disorder is implemented as deviations, specified in more detail below, from the perfect crystal symmetry.

The BTA morphology in the simulation has been adopted from what was discussed above. The hydrogen bonds forming between the dipolar amide groups have an out-of-plane rotation of 30° to 40° which leads to a helical pitch of six molecules with an interdiscular distance of 3.5 Å and an intercolumnar distance 17.2 Å for an alkyl chain with eight carbon atoms (BTA-$C_8$).[28–32] The only explicitly appearing part of the BTA structure are the amide groups in form of dipole vectors, while other parts are included implicitly, i.e. in lattice parameters such as intercolumnar distance (alkyl chain length) or distance between middle of a BTA molecule and the dipoles (size of benzene core). As opposed to molecular dynamics (MD) simulations, all particles are fixed in their position as given by the morphology throughout the simulation and the only mobile aspect is the orientation of the dipole vectors. Similar to an Ising model, the dipoles can take two states which are distinguished only by the z-component of the permanent dipole moment. As thin films are modelled, periodic boundary conditions in x- and y-directions are used; in the z-direction mirror boundary conditions are applied to account for the metal electrodes. They make use of the fact that the effect of a conducting plane on a charge $Q$ is equivalent to that of an oppositely charged mirror image $-Q$ on the other side of the electrode.

The final morphological aspect of disorder is implemented by separating the columns that would extend from electrode to electrode in an ideal situation into subcolumns of different lengths. The general simulated structure is columnar with the benzene cores of the BTA molecules stacking on top of each other until a defined height $n_z$ is reached. There are $x \cdot y$ columns which are simulated. The length for each subcolumn is drawn from a gaussian distribution with a mean value $N$ and variance $\sigma_{\mathrm{N}}$ and only integer values are allowed. The molecules within a column have a fixed rotation of $\alpha = 60°$ with respect to each other. To simulate defects, each subcolumn is separated from the others in the same stack by a shift $\sigma_{\mathrm{xyz}}$, a randomized rotation around the column axis and a potential change in helicity. At the start of a simulation run, the grid forming the BTA columns is formed with the specified parameters. An illustration of the different parameters and the symmetry breaking can be found in **Figure S1.1** in the Supplementary Information (SI). The parameters for a typical simulation are given in **Table S1.1** in the SI.

The flipping rates determining the evolution during the kMC simulation arise from the electrostatic interactions and the resulting energy calculations. The energy of a dipole $i$ is given by

$$U_i = -\vec{\mu}_i \cdot \vec{E}_i, \tag{2}$$

with the total dipole moment $\vec{\mu}_{\mathrm{i}}$ and the local electric field $\vec{E}_{\mathrm{i}}$. The total dipole moment consists of the permanent and induced contribution with the latter determined by the local field $E_{\mathrm{i}}$ and the polarizability $\alpha$ of the material:

$$\vec{\mu}_{\mathrm{i}} = \vec{\mu}_{\mathrm{i}}^{\mathrm{per}} + \vec{\mu}_{\mathrm{i}}^{\mathrm{ind}} = \vec{\mu}_{\mathrm{i}}^{\mathrm{per}} + \alpha \vec{E}_{\mathrm{i}}. \tag{3}$$

To calculate the energy for a dipole $i$, the local field at its position needs to be calculated. An externally applied electric field $\vec{E}_{\mathrm{app}}$ is assumed to be constant throughout the simulation box and its x- and y-components are chosen as zero. Then, the contribution to the electric field induced by and different dipole $j$ is given as

$$\vec{E}_{\mathrm{ij}} = \frac{1}{4\pi\epsilon_0\epsilon_{\mathrm{r}} r_{\mathrm{ij}}^3}\left(\vec{\mu}_j - 3\hat{r}_{\mathrm{ij}}\left(\vec{\mu}_{\mathrm{j}}\hat{r}_{\mathrm{ij}}\right)\right). \tag{4}$$

Here, $\epsilon_0\epsilon_{\mathrm{r}}$ is the material permittivity and $\vec{r}_{\mathrm{ij}} = r_{\mathrm{ij}}\hat{r}_{\mathrm{ij}}$ is the real-space vector (distance) separating the dipoles $i$ and $j$. Due to numerical limitations, the dipole-dipole interactions calculated this way need to be confined to a certain number of nearest neighbors. This defines a sphere with a cut-off radius $r_{\mathrm{c}} = 30$ inside of which all interactions are considered explicitly. The contribution of the remaining dipoles is not negligible due to the long-range nature of Coulomb interactions and is approximated using the reaction field method.[33] Within of this framework, it is assumed that all dipoles inside the sphere induce a polarization charge density $\vec{P}_{\mathrm{RF}} = \sum_{r<r_{\mathrm{c}}} \vec{\mu}_{\mathrm{i}}$ on the sphere surface and hence polarize the surrounding material. This leads to the following reaction-field at the center position:

$$\vec{E}_{\mathrm{RF}} = \frac{2(\epsilon_{\mathrm{r}} - 1)\vec{P}_{\mathrm{RF}}}{2\epsilon_{\mathrm{r}} + r_{\mathrm{c}}^{3}}. \tag{5}$$

Therefore, the complete local electric field can be written as

$$\vec{E}_{i} = \vec{E}_{\mathrm{app}} + \vec{E}_{\mathrm{RF}} + \sum_{\mathrm{j}, r_{\mathrm{ij}}<r_{\mathrm{c}}} \vec{E}_{\mathrm{ij}}. \tag{6}$$

Coupled with **Equation 2**, this provides the energy of dipole $i$. However, a self-consistency problem appears due to the feedback arising from the induced dipoles, as the induced dipoles are dependent on the local field which in return – as of **Equation 4 and 5** – depend on the induced dipoles. Thus, the problem is not solvable exactly and an iterative approach is chosen. Initially, the permanent dipole moment from the amide groups is taken as the total dipole moment. While the yielded values for the electric fields are naturally incorrect, they are used to calculate a better approximation of the actual dipole moment. Repeating these steps several times leads to converging values of $\vec{E}_{\mathrm{i}}$ and $\vec{\mu}_{\mathrm{i}}$ which are taken as the real values. Thus, the energies for all dipoles in any given dipole configuration can be obtained.

In particular, the energy difference $\Delta U_{\mathrm{i}}$ between a state with dipole $i$ up with respect to a state with dipole $i$ down, is the important factor for the flipping rate and is referenced to as the flipping energy. The value of $\Delta U_{\mathrm{i}}$ depends on the surrounding environment which changes significantly during the evolution of a simulation. Instead of solving the problem of the induced dipoles at each step of the kMC, an approximation of the change of the flipping energy is obtained beforehand. In addition to calculating $\Delta U_{\mathrm{i},1}$ for a given surrounding, $\Delta U_{\mathrm{i},2}$ is also obtained, where the neighboring dipole $j$ has been flipped. The difference of both yields $d_{\mathrm{j}}\Delta U_{\mathrm{i}}$ which is a measure of the effect of reversal of a neighboring dipole on the flipping energy of dipole $i$ and thus its flipping rate. Upon doing this calculation for all dipoles $j$ in the interaction sphere of dipole $i$, the flipping rate $\Delta U_{\mathrm{i}}$ can at all times throughout the evolution of one simulation be kept at the approximately correct value without recalculating the induced dipoles. Expanding this to all dipoles in the simulation box saves a significant amount of computational time.

The flipping energies are solely based on electrostatic interactions which depend on the morphology and can be calculated as

$$\nu_{\mathrm{i}} = \begin{cases} \nu_0 \exp\left(\frac{-\Delta U_{\mathrm{i}}}{k_{\mathrm{B}}T}\right) & \text{if } \Delta U_{\mathrm{i}} > 0 \\ \nu_0 & \text{otherwise.} \end{cases} \quad (7)$$

Here, $\nu_0$ is the attempt-to-flip frequency which is taken as a typical vibrational frequency of the material and $k_{\mathrm{B}}T$ is the thermal energy in the system. During the kMC simulation, the flipping rates serve as weight factors in determining which dipole flips. The probability for a dipole $i$ to be selected for a flip is given by $\nu_{\mathrm{i}}/\nu_{\mathrm{tot}}$ with the sum of all flipping rates $\nu_{\mathrm{tot}} = \sum_i \nu_{\mathrm{i}}$. At each simulation step, a random number between zero and one is drawn which decides which dipole is flipped. A second random number $r_2 \in [0,1]$ is used to calculate the increment of time associated with the current step:

$$\Delta t_{\mathrm{i}} = \frac{\log(r_2)}{\nu_{\mathrm{tot}}}. \quad (8)$$

This ensures that in parts of the simulations where many dipoles are volatile and thus $\nu_{\mathrm{tot}}$ is large, each flip on average is much faster than in a situation where the vast majority of dipoles is stable. After each kMC step, all flipping energies are updated based on the energy calculations done before the simulation. Saving the state of all dipoles at each time step and converting it to a physical polarization provides a close resemblance of the numerical simulations to a real experiment. Since the external field is a free parameter, hysteresis loops and switching transients can be recorded in addition to simple (de)polarization studies.

## 3. Simulation and analysis procedure

In all numerical experiments, the BTA system was driven with a slowly varying external field while measuring the polarization. All dipoles initially pointed downwards, hence providing a fully saturated starting condition. The parameter representing the electric field was ramped from -0.8 V $\mathrm{nm}^{-1}$ to 0.8 V $\mathrm{nm}^{-1}$ and back again. The complete driving was spread over 1 microsecond and 120 different field values where each field value represents a sub-simulation with recalculated induced dipole problem. Typical simulation settings are provided in Table S1.1 in the SI.

**Figure 2a** shows a typical result for a simulated switching process over time with the driving field $E$ (blue), the permanent polarization $P_{\mathrm{p}}$ (orange) and the obtained current of a complete loop $I$ (green), i.e. the concatenation of all sub-simulations. The current represents the number of flipped dipoles per time bin and bundled up over longer time frames as jerks. Large jumps in the permanent polarization and spikes in the current are observed once the applied electric field is large enough, i.e. approaches the coercive field of the simulated system. Some of the spikes are numerical artefacts as seen by their regular occurrence and are due to the abrupt changes of the applied field and the corresponding recalculation of the linear polarization and flipping rates. Filtering out those data points leads to the curves of total polarization shown in **Figure 2b**, where the majority of the current peaks remains, indicating the presence of crackling noise.

To identify the individual events, a threshold level was set, as is indicated in Figure 2b (horizontal black lines). This is required since dipole flips are often initiated due to thermal fluctuations, although they reverse shortly afterwards due to the local field still pointing in the other direction. While many relatively small events arise this way, they do not contribute to the switching process and have to be filtered out. An event starts when the absolute of the polarization current first exceeds the threshold level and ends once it falls below it again. As an alternative to the polarization current, events can be defined by counting the dipoles that flipped per time step which is the preferred method as it offers an immediate measure of the avalanche sizes while the polarization current depends on the sample geometry. However, these two quantities are linearly proportional, hence the current is the obvious choice as a proxy for the avalanche sizes in a real experiment.

Setting the threshold level is a trade-off between losing small crackling events and adding events that are thermal noise generated, as illustrated in **Figure 2c** showing the number of detected events as a function of the threshold level normalized to the largest measured spike $x_{\mathrm{max}}$ of one loop. Furthermore, as the typical number of events collected from a single loop is insufficient for statistical analysis, multiple single numerical experiments were combined, as shown in **Figure 2d**. The full description of the threshold settings, the analysis of simulated events and combination of similar data sets is provided in Section S2 in the SI.

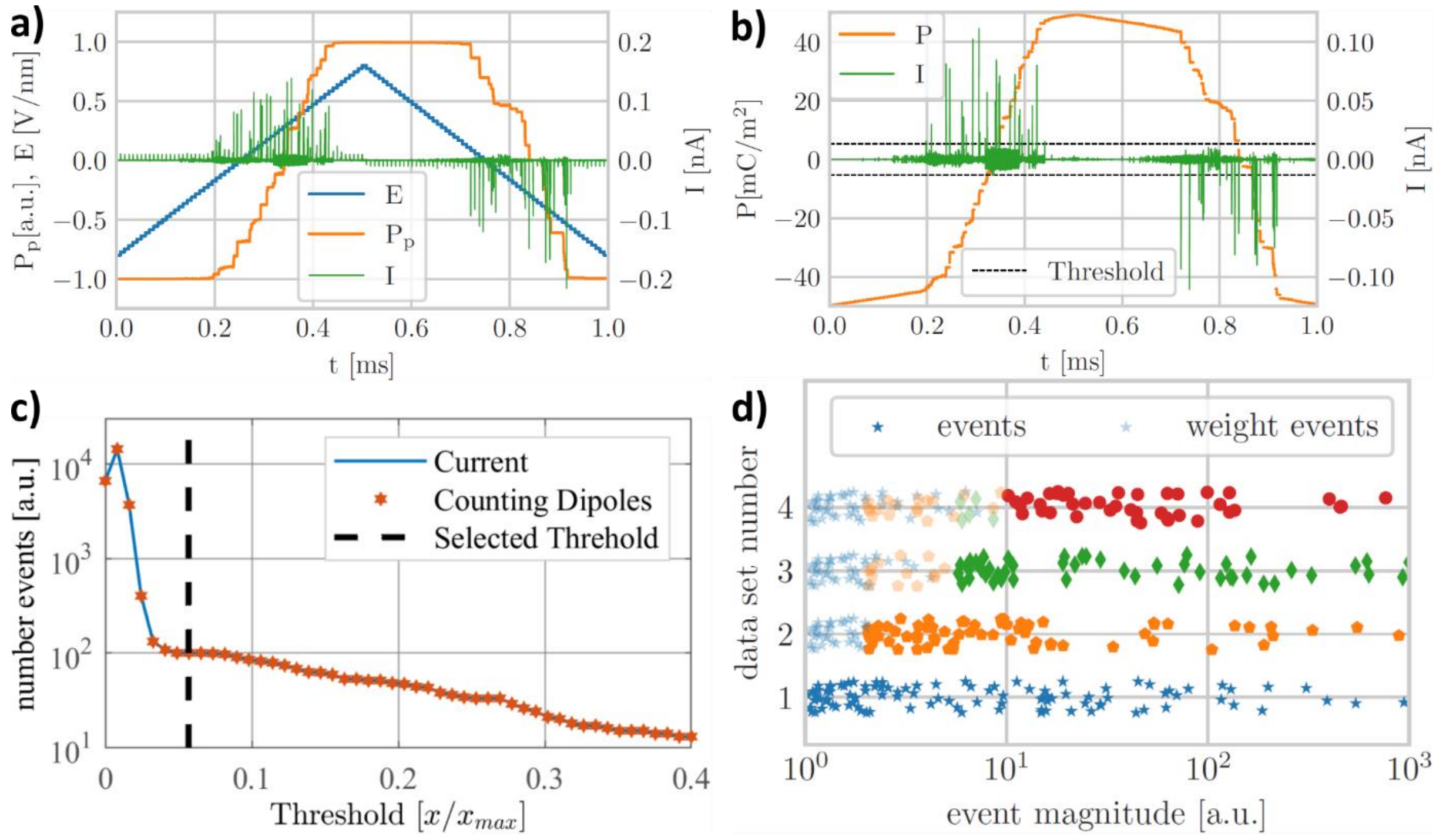

**Figure 2:** Typical results from a numerical experiment **a)** before and **b)** after artefacts due to the abruptly changing field are filtered. The current shows significant spikes that constitute dipole avalanches. A threshold level (black horizontal dotted lines in b) is applied to identify single events and is chosen to be just above the thermal background noise. **c)** The reduction of the threshold level leads to an increase of the number of events detected for a single simulated loop. The optimal threshold level is set to be just beyond the kink due to the inflated number of small (thermal) events due to thermal fluctuations (black vertical line in c). **d)** Upon merging multiple individual simulations (denoted by different symbols) with different threshold values, a correction is needed in the regime of small events. It was implemented by creating weight events based on those data sets that extent to this regime. The simulation parameters here were $f = 1$ kHz, $n_\mathrm{x} = n_\mathrm{y} = 4, n_\mathrm{z} = 300, N = 10, \sigma_\mathrm{N} = 1, T = 300$ K.

The results of an exemplary set of simulations and the corresponding probability densities are shown in **Figure 3**. Here, the size $S$ is defined as he highest number of flipped dipoles per time step during one avalanche and the summed size $S_\Sigma$ is obtained by adding up all dipoles that flipped per time step in the course of one event. The latter was not further analyzed as it is subjected to artifacts arising from large fraction of singular events. The energy of an avalanche $U$ is defined via its proportionality to the integral of the electrical power and the inter-event time $\delta t$ is the time-span between events. The probability density functions for the sizes and energies show a similar structure with an initial region of small events that are not captured by the power-law, followed by a large part of the distribution showing the power-law behavior. The distribution shows a steep decline for the biggest events due to the limitations arising from the system size. For size $S$ and summed size $S_\Sigma$ distributions, a clear shift towards larger events as a result of summing over all points within an event is visible. The energy distribution shows same trend as the distribution of event sizes and the proportionality $U \propto I^2 \propto S^2$ is clearly visible from the spread of the values. The inter-event times are directly tied to the time given by the sweeping frequency of the applied electric field. Here, the simulated time was $t = 10^{-3}$ s and all the simulated events happened within the time frame of $\delta t \approx 2 \cdot 10^{-4}$ s.

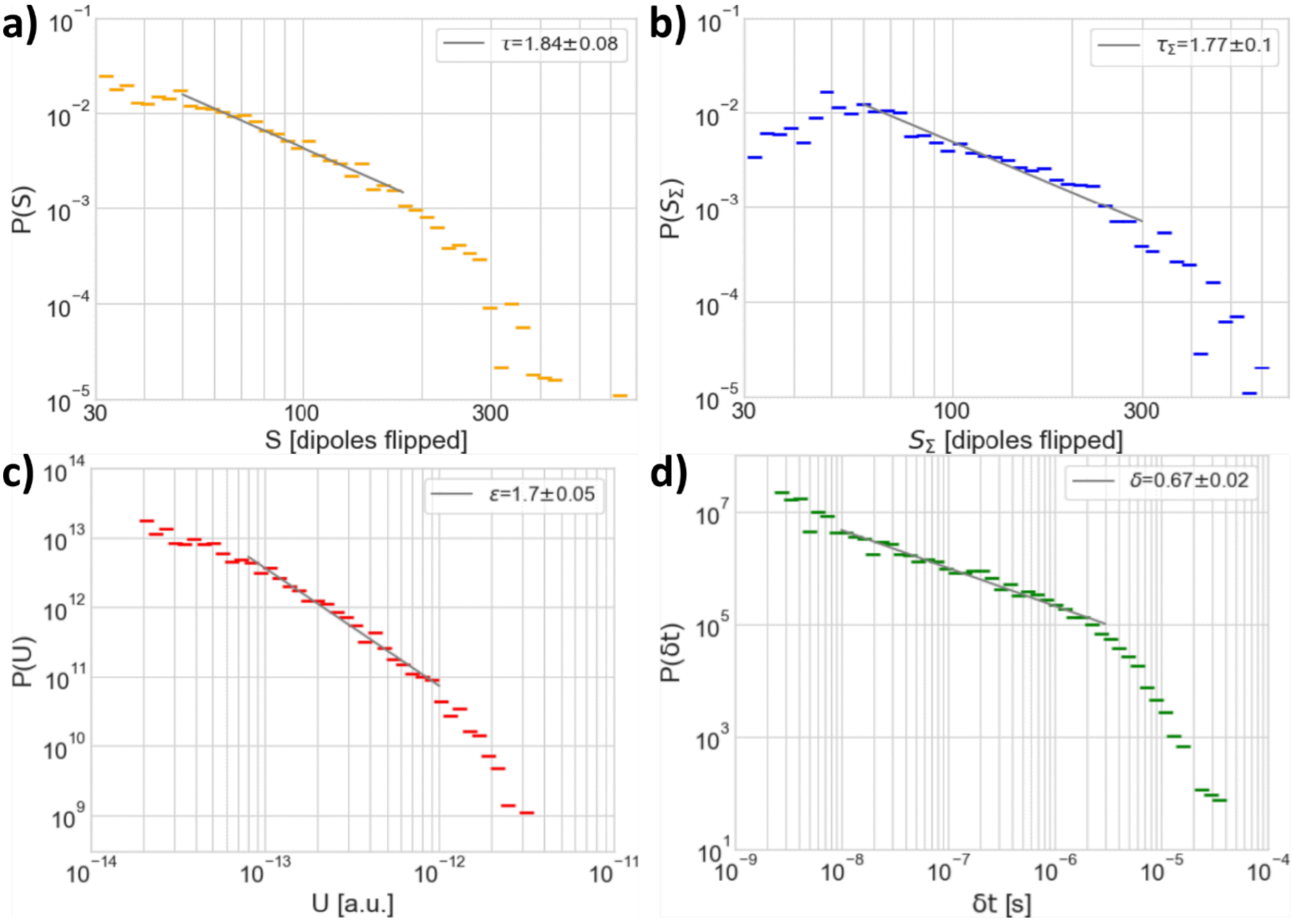

**Figure 3:** Probability densities for **a)** size, **b)** summed size, **c)** event energy and **d)** inter-event time. All distributions exhibit similar structure with a region showing a power-law behavior (fitted data) followed by a sudden decline due to the limitations of the system size or, in case of inter-event time, simulation duration. The simulation parameters here were $f = 1$ kHz, $n_x = n_y = 4, n_z = 300, N = 10, \sigma_N = 1, T = 300$ K, number of events $= 1911$.

## 4. Simulation results

A general concern with the numerical simulation of potentially critical behavior of finite systems is that a relatively higher number of larger events could happen in bigger simulation boxes. To assure that finite size effects do not affect our results, additional simulations with gradually increased system size $n_z$ were carried out. **Figure S3.1** in the SI shows the probability densities for six different sizes between 40 and 700 molecules in the z-direction. It was observed that the occurrence of larger event sizes increased with increasing column length until $n_z \approx 300$. For even larger systems, barely an effect was found, hence the main simulations were carried out with a system size of $n_z = 300$. We note that this corresponds to a film thickness of $\sim 105$ nm, which is less than the film thicknesses typically investigated experimentally.

Furthermore, to investigate the effect of the column height on the event sizes, the mean of the event sizes was analyzed for different system sizes. A strong initial increase which diminishes with the increase of the system size was observed, as depicted in **Figure 4a**. The same effect was also investigated for systems with a larger number of shorter columns in the x- and y-directions, which is shown in **Figure 4b**. The resulting distribution showed generally smaller events. Additionally, the effect of adding more columns while keeping their length constant was investigated, which is presented in **Figure S3.2** in the SI. As no limitations due to the geometry are apparent in contrast to varying the box height (see Figure 4a), one dimensional avalanche propagation along the column axis is recognized as the dominant mode of switching. This behavior is attributed to the morphology, in particular the large difference between hexagonal packing distance and interdisc distance, as the dipoles are closer packed along the z-direction and hence an initial thermal perturbation is also more likely to be communicated along that direction, thus leading to needle-like dipole avalanches.

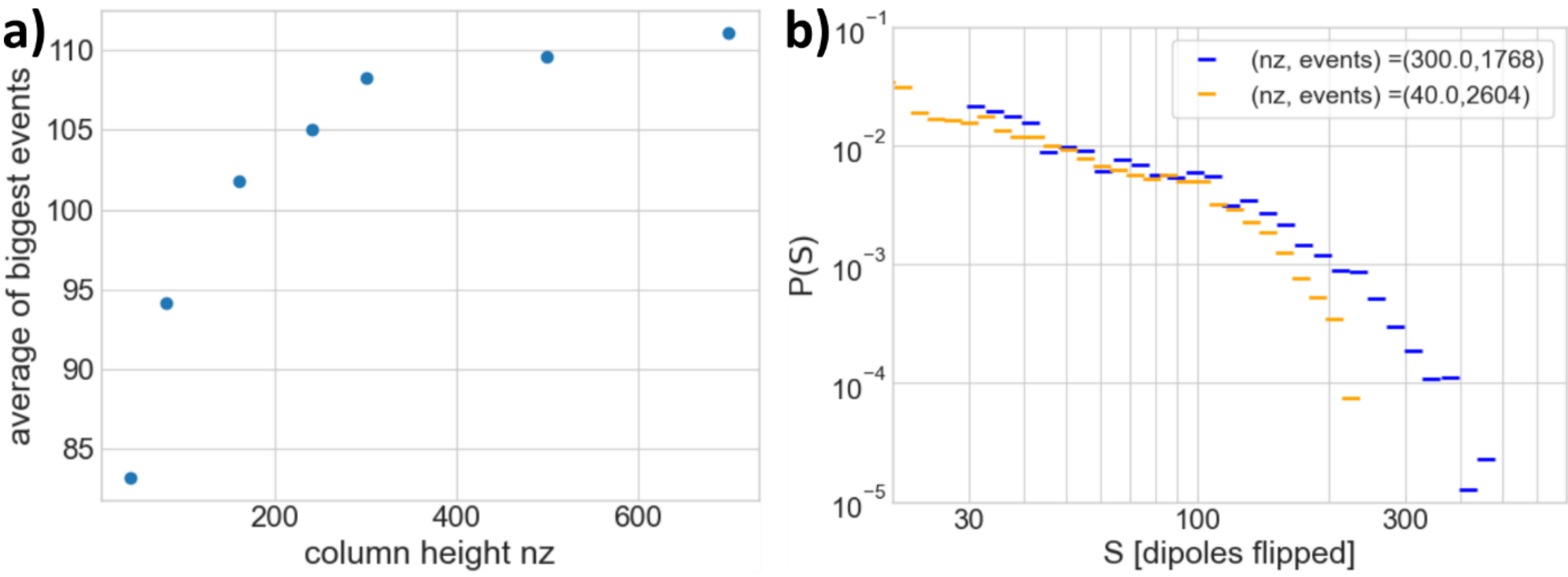

**Figure 4:** Effects of different sizes within the simulated systems. The simulation parameters here were $f = 1$ kHz, $N = 10, \sigma_N = 1, T = 300$ K. **a)** The mean value of events surpassing the threshold $S = 40$ for different column lengths increases with higher column heights $n_z$,

although the effect strongly diminishes for higher values ($n_z > 300$). **b)** Comparison of a distribution with large column size but fewer columns $\left(n_x = n_y = 4, n_z = 300\right)$ to one with a similar number of dipoles with increased number of smaller columns $\left(n_x = n_y = 11, n_z = 40\right)$. Despite an increase in the lateral system size, only the increase in column heights leads to a significant increase of larger events.

In summary, two conclusions can be drawn here. First, dipole avalanches propagate almost exclusively along the column axis. Second, for the parameter settings used above, the size of the largest avalanche was found to not be limited by the height of the system, provided the systems thickness exceeds $\sim 100$ nm. Combined, this puts a rather stringent upper limit on the maximum event size that occur, in turn implying that truly critical behavior, which is scale-invariant over all investigated length scales, does not occur at the used temperature $T = 300$ K. In the following, we will see that critical behavior does occur at lower temperatures.

Having established the occurrence of jerky behavior in the polarization switching of BTA and of power-law distributions of event characteristics, we now focus on the specific values of these exponents and their dependence on temperature *T*, sweeping frequency *f* and material characteristics. A brief discussion on the role of temperature *T* and sweeping frequency *f* can be found in **Section S4** in the SI. First, we focus on the role of temperature, that was swept between 100 and 300 K. As expected for ferroelectric materials, lower temperatures require higher fields to start the switching process and – depending on the chosen applied electric field – not the whole simulation box may be switched for a constant maximum field at the lowest temperatures. A comparison of typical simulations for 100 and 250 K is shown in **Figure S4.1** in the SI.

A comparison of the corresponding distributions is displayed in **Figure 5**, showing a shift to bigger event sizes for lower temperatures in both the size and energy distributions. This is expected as fewer thermal excitations lead to a higher electric field required for switching and thus one event is more probable to cause bigger avalanches. This leads to the exponents recovered from the power-law to be significantly smaller for the simulation at 100 K.

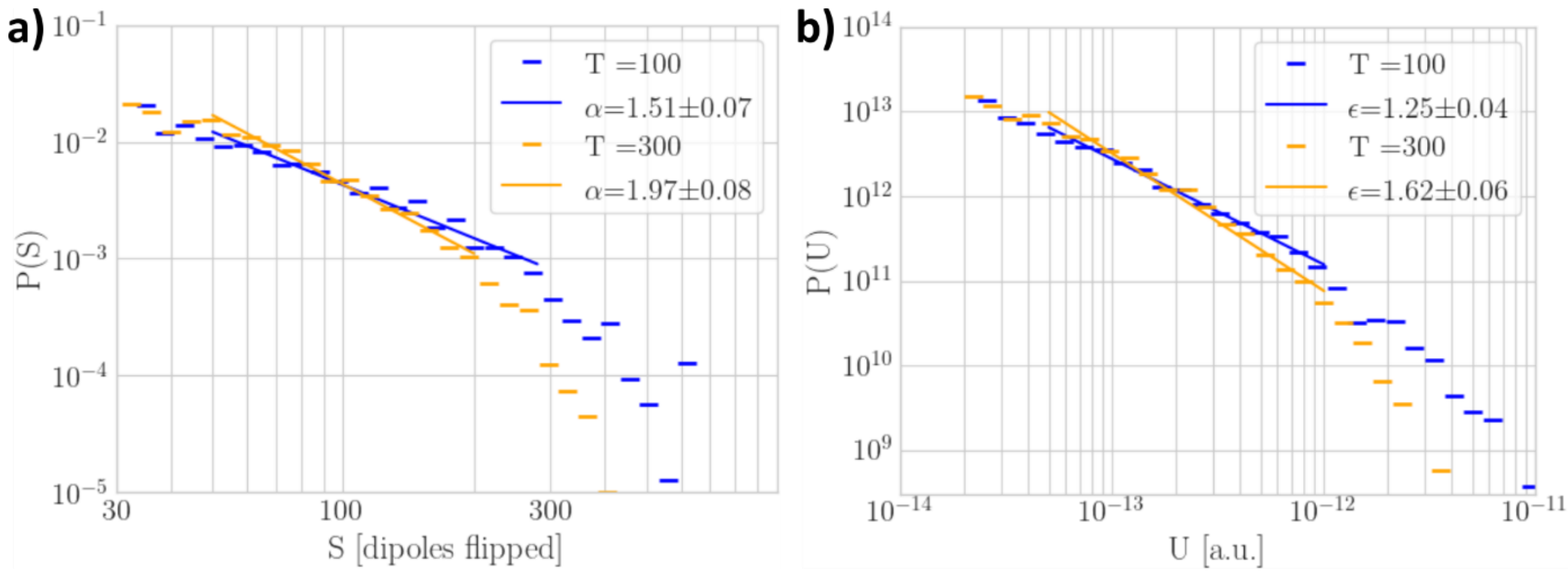

**Figure 5:** Probability distribution of **a)** event size and **b)** energy for the same system settings at a lower ($T = 100$ K) and higher ($T = 300$ K) temperature. The simulation parameters here were $f = 1$ kHz, $n_x = n_y = 4, n_z = 300, N = 10, \sigma_N = 1$.

The results of the systematic temperature sweep are displayed in **Figure 6a**. For the used (experimentally realistic) parameters, temperatures below 100 K would often result in only very little switching throughout the process, whereas above 300 K, the thermal noise present in the system led to large variations in the datasets which are already visible for the highest temperatures shown. Both the size and energy distributions show a similar trend for the entire temperature range. At low temperatures, the extracted exponents are consistent with mean field predictions of system-spanning avalanches which are indicated by horizontal lines. The agreement of the numerical values with those predicted from theory in combination with their relatively constancy at lower temperatures, suggest that the field-driven switching process in BTA shows self-organized criticality at temperatures below ~170 K.

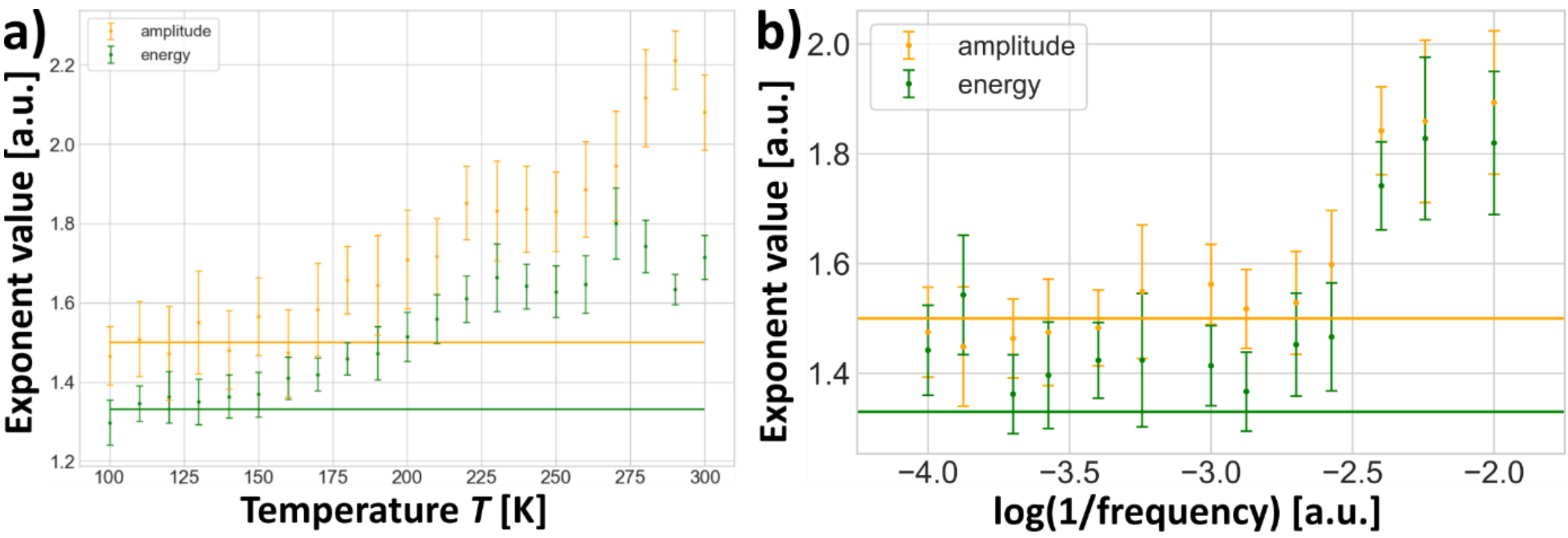

**Figure 6:** Size (orange) and energy (green) power-law exponents extracted from simulations of bigger systems for **a)** different temperatures at $f = 1$ kHz and **b)** sweeping frequencies at $T = 160$ °C. The horizontal lines indicate the values calculated from the mean field plasticity model. The simulation parameters here were $n_\mathrm{x} = n_\mathrm{y} = 4, n_\mathrm{z} = 300, N = 10, \sigma_\mathrm{N} = 1$.

With increasing temperature, the exponents increase, implying that the system favors smaller, localized events which leads to a deviation from the mean-field prediction and indicates the system enters a creep-regime: as the thermal energy present in the system increases, purely field-induced events are more likely to get replaced by smaller and (partially) thermally induced events. In contrast, at lower temperatures, the switching happens at higher electric fields, which increases the probability for the flipping of one subcolumn to cause further events in one avalanche, leading to bigger events having a higher probability and, ultimately, self-organized criticality.

Furthermore, we investigated the effect of the sweeping frequency $f$ on the power-law exponents, as shown in **Figure 6b**. As we established the self-organized criticality in BTA below 170 °C, the sweeping frequency was varied at $T = 160$ °C. Both the size and energy exponents are consistent with the mean field prediction at higher sweeping frequencies and increase with decreasing sweeping frequencies at lower investigated sweeping frequencies. It is apparent that the power-law exponent trends for T and $\log(1/f)$ are similar.

A probable explanation of the similar trends is related to the thermally activated nature of switching processes in organic ferroelectrics. The thermally activated nucleation limited switching (TA-NLS) model assumes switching to be limited by the activation of pre-existing nucleation sites in the small regions of which the material consists of.[34,35] The activation of a nucleation site is described as the thermally driven polarization reversal of a critical volume $V^*$. The model yields the following expression for the coercive field $E_\mathrm{c}$ dependent on the voltage sweeping frequency $\nu_\mathrm{exp}$:

$$E_\mathrm{c} = \frac{w_\mathrm{b}}{P_\mathrm{s}} - \frac{k_\mathrm{B}T}{V^* P_\mathrm{s}} \ln\left(\frac{\nu_0}{\nu_\mathrm{exp} \ln 2}\right). \quad (9)$$

Here, $w_\mathrm{b}$ is the energy barrier between two metastable states, $P_\mathrm{s}$ is the saturation polarization, $k_\mathrm{B}T$ the thermal energy in the system and $\nu_0$ is the attempt-to-flip frequency, which is associated with a material-specific phonon frequency. A high sweeping frequency in a hysteresis loop measurement leads to a large coercive field, as the system does not have enough time to equilibrate and relax to the state of its lowest energy.

The TA-NLS model predicts that the coercivity of a ferroelectric system grows as the driving frequency increases and lower driving frequencies accordingly make thermal fluctuations more pronounced. **Equation 9** shows that the coercive field decreases linearly with temperature. The logarithm of the inverse frequency has the same effect. Based on Figure 6, a strong similarity of the trends for power-law exponents as functions of temperature and logarithm of the inverse frequency is apparent. Therefore, we conclude that the power-law exponents are anti-correlated to the coercivity of the system.

If indeed the chiral columnar BTA system shows self-organized criticality upon field-driven polarization reversal at lower temperatures, the observed exponents should be invariant to modest changes in system parameters. In order to verify that, further simulations were performed in the low temperature range where the critical exponents stay consistently around the predicted mean field values. The investigated disorder parameters included the variance of the distribution from which the subcolumn sizes were generated and the chirality of the defects.

Investigating the different types of defects, the obtained result was that having a single type of defect can lead to either stabilization or destabilization of the system, depending on the geometrical details of the defect and the corresponding deviations in exponents. Qualitatively spoken, a 'sufficient' variation in defects is needed to obtain universal exponents. Specifically, the introduced disorder needs to be big enough to prevent completely intrinsic switching whereas too much disorder in the system leads to small events dominating the switching process. This behavior is consistent with the earlier mentioned Random-Field-Ising model.[36] For the present system, this implies that both heterogeneity in defect type as well as in subcolumn length are necessary conditions to recover the universal exponents. The details of the investigation are discussed in **Section S5** in the SI.

## 5. Experimental results

To investigate whether the simulated results could be experimentally confirmed, measurements on both out-of-plane and in-plane BTA-$C_{10}$ samples were conducted. The biggest challenge was to reduce the noise level of the measurement setup while keeping sufficient time resolution as the expected currents associated with single jerks are very small. The finalized measuring setup is described in the experimental details section. The measured noise floor is depicted in **Figure S6.1** in the SI and the corresponding noise levels are given in **Table S6.1** in the SI.

Despite the carried-out noise reduction we did not succeed in measuring Barkhausen noise in the investigated BTA-$C_{10}$ samples. Typically measured current transients and the corresponding obtained probability densities are presented in **Figure S6.2 and S6.3** in the SI, respectively.

Also, in view of the fact that we successfully measured Barkhausen noise in P(VDF-TrFE),[37] this raises the suspicion that the switching events are too small in BTAs. To verify this, we estimate the number of dipoles flipped within our samples and compare this number to both the simulations and noise level of the setup. The corresponding calculation can be found in **Section S7** in the SI. As of our findings, the number of regions that would need to be simultaneously switched in order to achieve a measurable current is far larger than what is plausible on basis of the weak inter-columnar coupling. In other words, the negative result of the experiments, i.e. Barkhausen noise being below the resolution threshold of the used setup, is consistent with the predictions of the numerical model.

## 6. Summary

Barkhausen-noise in the columnar hexagonal liquid crystalline ferroelectric BTA was investigated. Numerical simulations by kinetic Monte Carlo show that the material exhibits critical behavior over long ranges of the event sizes at temperatures $T \lesssim 175$ K for a wide range of disorder parameters, while a minimum amount (but not too much) of the latter is required. The disorder present in the system was varied by changing the subcolumn size and its variation, which corresponds to the defect density, as well as by varying the nature of the defect. The resulting behavior is similar to that of a random field Ising model. The behavior of the system and changes in power-law exponents were investigated for a wide range of temperatures. At lower temperatures ($\lesssim 175$ K), and equivalently higher driving frequencies, the extracted power-law exponents were found to be similar to the mean field predictions for self-organized critical behavior, while higher temperatures and lower driving frequencies led to systematically larger values, indicating thermal creep, i.e. many switching events originating from thermal excitations rather than the driving field. While the critical power-law exponents were affected to at least some degree for all examined parameters, the changes were significant only outside a certain regime (identified as the self-organized critical regime), hence justifying the assumption of self-organized criticality in the simulated system. Measuring Barkhausen noise of BTA-$C_{10}$ experimentally was unsuccessful due to too small current signals produced by the material during the switching process, in agreement with the event sizes expected on basis of the numerical simulations.

## 7. Experimental details

**Measuring setup:** The input signal was supplied by a Keysight 33600A Arbitrary Waveform Generator (AFG) and was amplified by a TReK PZD350A high voltage amplifier. Multiple AFGs (Keysight 33600A, Tektronix AFG3052C and Tektronix AFG1062) and amplifiers (TReK PZD350A and Falco System WMA-200) were tested and these were to be found to have the lowest noise level, although the difference was marginal, as shown for AFGs in **Figure S6.4** in the SI. The device response was measured by either a Zurich Instruments impedance analyzer (MFIA) or Lock-In amplifier (MFLI). No significant differences between the MFIA and MFLI were established and an oscilloscope could be used instead given it has sufficient resolution. The device under test was measured inside a Linkam stage in order to reduce noise and allow for precise temperature dependent measurements.

**Sample preparation:** Both out-of-plane and in-plane electrodes were used for device fabrication of thin film BTA samples. In case of out-of-plane samples, commercially available Corning plain microscope slides with a thickness of 0.96 to 1.06 mm were cut and both bottom and top electrodes were thermally evaporated. Both aluminum on silver and gold on chromium electrodes of different thicknesses were utilized. The BTA thin films were spin coated with 900 rpm. As in-plane electrodes, interdigitated electrodes (IDEs) supplied by MicruX Technologies were used. Here, the material was drop casted from solution.

## Conflicts of interest

The authors declare no conflicts of interest.

## Data availability

The data supporting this article have been included as part of the manuscript and its Supplementary Information.

## Acknowledgements

We thank the Deutsche Forschungsgemeinschaft (DFG, German Research Foundation) for support of this work (SFB 1249 and EXC-2082/1-390761711). M.K. thanks the Carl Zeiss Foundation for financial support.

Supporting information to

**Barkhausen noise in the columnar hexagonal organic ferroelectric BTA**

*Andrey Alekseevich Butkevich, Fabian T. Thome, Toni Seiler, Marcel Hecker, Martijn Kemerink*

Andrey Alekseevich Butkevich, Fabian T. Thome, Toni Seiler, Marcel Hecker, Martijn Kemerink

Institute for Molecular Systems Engineering and Advanced Materials, Im Neuenheimer Feld 225, Heidelberg, 69120, Germany.

Email: martijn.kemerink@uni-heidelberg.de

## S1. Details on simulated parameters and material structure

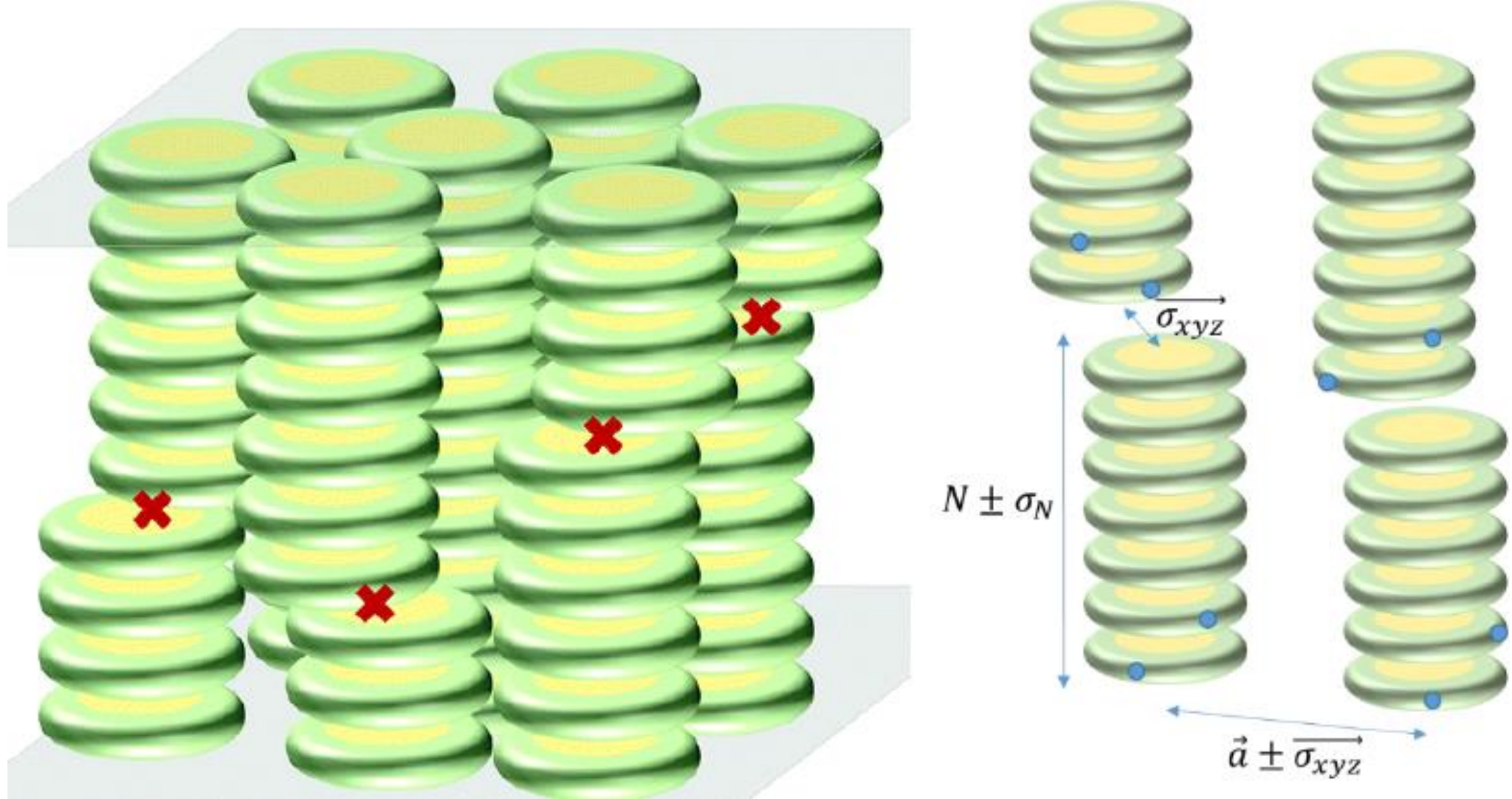

**Figure S1.1:** Columnar structure of BTA showcasing the displacements due to defects (red crosses) and different length scales. Within a BTA discotic, the benzene core and amide group spacing are represented in yellow and green, respectively. Defects separate the BTA columns spanning from one electrode to the other into subcolumns of length *N* and variance $\sigma_{\text{N}}$. Each defect consists of a shift in space $\sigma_{\text{xyz}}$ with respect to the original position as defined by the crystal geometry, a rotation around the column axis and a possible change in helicity.

| | Parameter | Value | Unit |
|---|---|---|---|
| Morphology | Helical pitch | 6 | molecules |
| | Dipole out-of-plane rotation β | 40 | degrees |
| | Hexagonal packing distance $a$ | 1.7 | nm |
| | Columnar interdisc distance $c$ | 0.35 | nm |
| | Dipole distance from center $L$ | 0.28 | nm |
| Disorder | Subcolumn length $N$ | 2 – 100 | molecules |
| | Subcolumn length variance $\sigma_{\mathrm{N}}$ | 0 – 50 | molecules |
| | Positional shift $\sigma_{\mathrm{xyz}}$ | 0.01 | nm |
| Simulation | Driving frequency $f$ | $10^2 - 10^5$ | Hz |
| | Box size in x-direction $n_{\mathrm{x}}$ | 4 – 11 | molecules |
| | Box size in y-direction $n_{\mathrm{y}}$ | 4 – 11 | molecules |
| | Box size in z-direction $n_{\mathrm{z}}$ | 40 – 700 | molecules |
| | Interaction cut-off range $r_{\mathrm{c}}$ | 30 | dipoles |
| | Total number of dipoles $n$ | $n = 3n_{\mathrm{x}} \cdot n_{\mathrm{y}} \cdot n_{\mathrm{z}}$ | dipoles |
| Materials | Permanent dipole moment $\mu_{\mathrm{per}}$ | 4 | Debye |
| | Polarizability $\alpha$ | 1 | e Å$^2$ V$^{-1}$ |
| | Temperature $T$ | 100 – 375 | Kelvin |
| | Attempt-to-flip frequency $\nu_0$ | 1 | THz |
| | Effective permittivity $\varepsilon_{\mathrm{r}}$ | 2 | |

**Table S1.1:** Simulation parameters typically used. The subcolumn length $N$ is a measure of the disorder and the higher the value the smaller amount of the disorder in the system. Parameters that are given a single value were set constant for all simulations.

## S2. Details on the threshold settings

Setting the threshold level is a trade-off between losing small crackling events and adding events that are thermal noise generated. Figure 2c in the main text shows the number of detected events as a function of the threshold level normalized to the largest measured spike $x_{\mathrm{max}}$ of one loop for both methods. The identical behavior of the curves shows the equivalence of each approach. Here, the threshold value was determined by setting it manually on the plateau just right of the thermal noise peak as this position corresponds to the optimal trade-off between excluding thermal fluctuations and including irreversible dipole avalanches and its validity is further confirmed by inspecting Figure 2b in the main text. The value of the threshold fluctuated at around 15 dipoles per time step. In addition to collecting data from the complete loop, events exclusively arising from the switching activity close to the time where the polarization is zero can be analyzed separately which corresponds to investigating the system at the coercive field only. The approach allows for a differentiation between field-specific and field-integrated event distributions.

After setting the threshold level, the events were analyzed for their characteristic properties. The duration of an event was given by the time difference of the intersection points between threshold and switched dipoles (or current). In the simulations, an event was typically completed within a few bins, thus rendering a statistical analysis of their width in terms of PDFs nonsensical. This is attributed to a combination of finite size effects and the tendency of the kMC towards intrinsic switching.[1] The remaining investigated characteristics generally showed sufficiently broad distributions.

As established previously, distributions of crackling quantities commonly follow power-laws. More precisely, power-law distributed means that the probability of finding an event with value $X$ in the interval $[x, x + dx)$ is given by

$$p(x)dx = P(x \leq X < x + dx) = Cx^{-\alpha}dx \qquad (S2.1)$$

with the normalization constant $C$. In practice, the power-law holds only above some finite minimal cut-off value $x_{\mathrm{min}}$, as otherwise the normalization condition

$$1 = \int_{x_{\mathrm{min}}}^{x_{\mathrm{max}}} p(x)dx \qquad (S2.2)$$

could not be met. In real systems, a finite upper cut-off $x_{\mathrm{max}}$ is also expected as truly infinite events cannot occur. The range in which the power-law would be fitted was determined by visualizing the event data. Bins that increase logarithmically in width along the event axis were used to reduce the effect of significant fluctuations in the tail of the distribution. The computed probability was afterwards divided by the width of the bin. The complete probability density function was normalized as of **Equation S2.2**. To get stable fit results, both the probability density $p(x)$ and the bin positions were log-transformed and a linear least square fit was executed. After reversing the transformation, the power-law fit was extracted. This method was compared to the so-called maximum likelihood estimation that is widely used in crackling noise studies on ferroelectrics.[2–5]

Since the typical number of events collected from a single loop was generally insufficient for a meaningful statistical analysis, the data collected from multiple single numerical experiments with equal settings was combined. This led to the implementation of a correction involving creating weight events as duplicates of the simulation results reaching the lowest limit in order to circumvent the issue of merging data sets with potentially different threshold levels, as the combined data featured an artificially reduced probability density for the (summed) size and energy in the regime of small events. The approach is illustrated in Figure 1d in the main text for four synthetic data sets with different threshold levels. A similar correction for interevent times was not required as those are not correlated to the event size or energy.

## S3. Effects of simulated system size

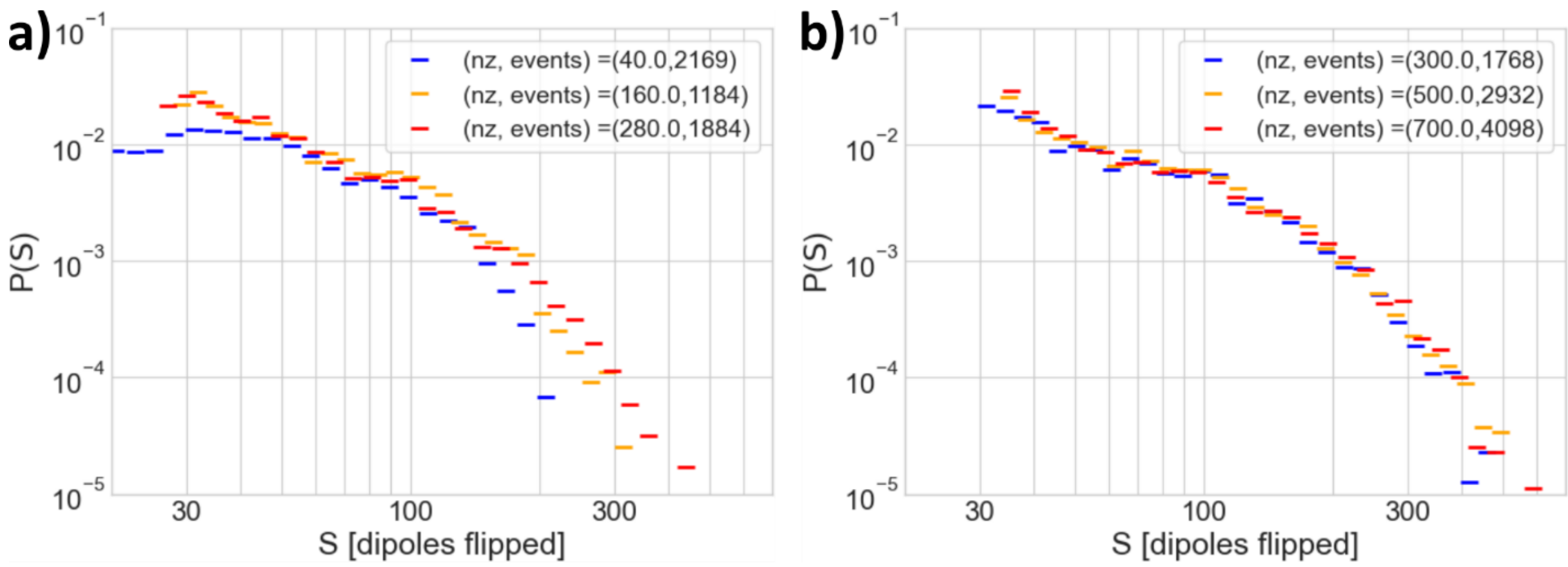

**Figure S3.1:** Probability densities for different sizes of the simulated system. **a)** An increase in the number of bigger switching events is observed with increasing size until $n_z \approx 300$. **b)** For even larger system sizes, this effect strongly decreases. The simulation parameters here were $f = 1$ kHz, $n_x = n_y = 4, \sigma_N = 1, N = 10, T = 300$ K.

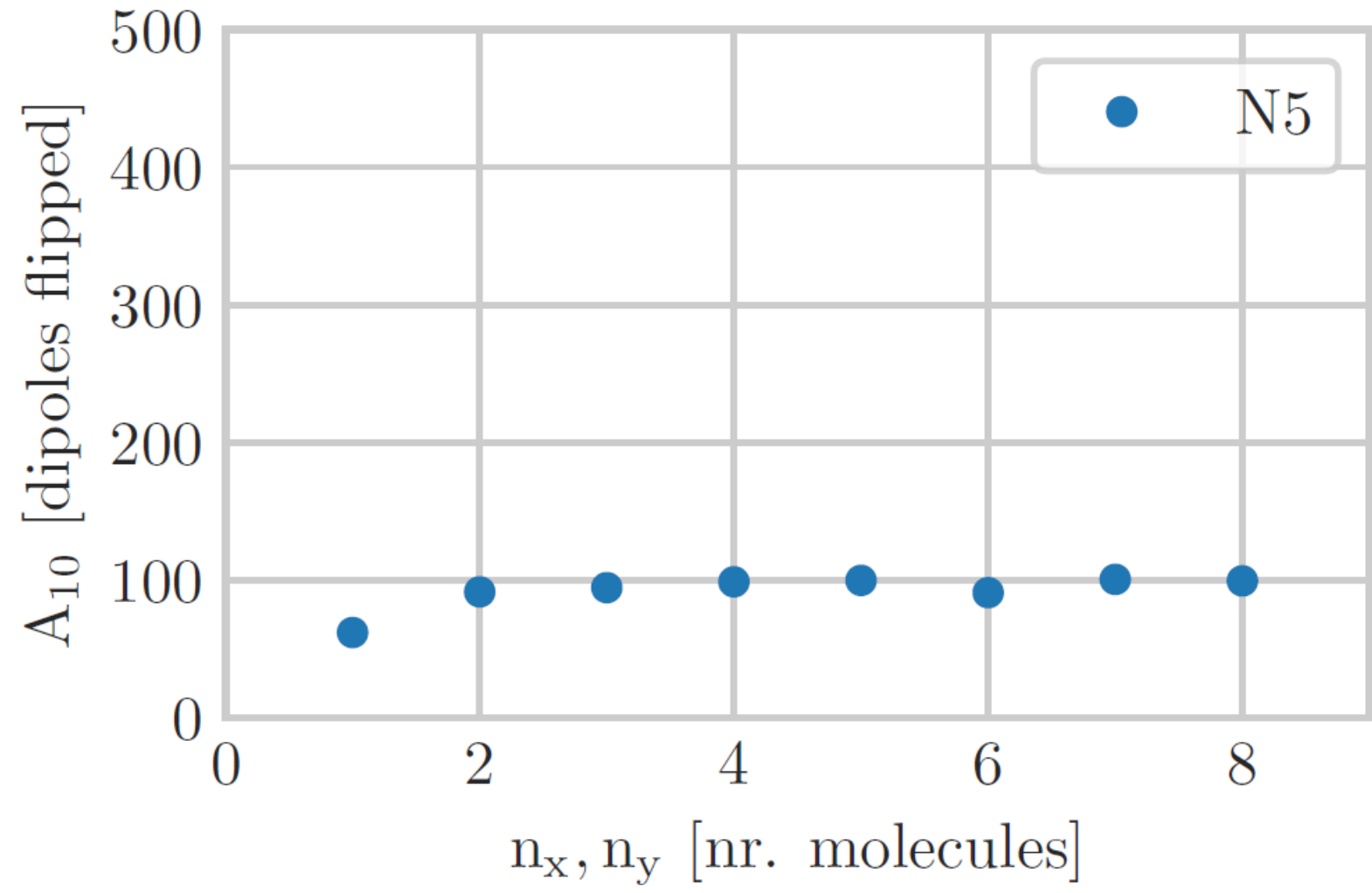

**Figure S3.2:** The median of the ten largest simulated dipole avalanches for different box widths $(n_x = n_y)$ and a box height of 20 molecules. Once more than a single column is included, the box width has no apparent influence on the size of the largest avalanches, hence indicating one dimensional avalanche propagation. The simulation parameters here were $f = 1$ kHz, $n_z = 20, N = 5, \sigma_N = 1, T = 300$ K.

### S4. Effects of temperature and sweeping frequency

Since switching in organic ferroelectrics is a thermally activated process, the temperature is one of the investigated parameters. Figure 6a in the main text shows the power-law exponents for temperatures ranging from 100 to 300 K. Both the size and energy exponents have an initial plateau until $T = 160$ °C and are clearly increasing with temperature above that. Higher power-law exponents indicate that the switching is more focused on smaller events. As $k_\mathrm{B}T$ increases, the switching becomes more volatile and small events where the system jumps between neighboring meta-stable states occur more often. The role of larger avalanches diminishes and a tendency to slow and steady spreading of the polarization reversal arises which is known as the creep regime and has been experimentally confirmed to lead to higher power-law exponents.[6] In total, the values of the critical power-law exponents are subject to temperature which is attributed to the onset of the thermal creep regime. In all observed cases, the critical character of the switching was not touched as system-spanning avalanches still appeared side by side with smaller ones.

Furthermore, the driving frequency was investigated in the range from $10^2$ to $10^4$ Hz for a temperature of $T = 160$ K. For these simulations, the bin-size was fixed at 1 ns. Figure 6b in the main text shows the power-law exponents extracted from the size and energy PDFs as a function of the logarithm of the inverse frequency. Generally, both the size and energy power-law exponents follow the same trend as for temperature.

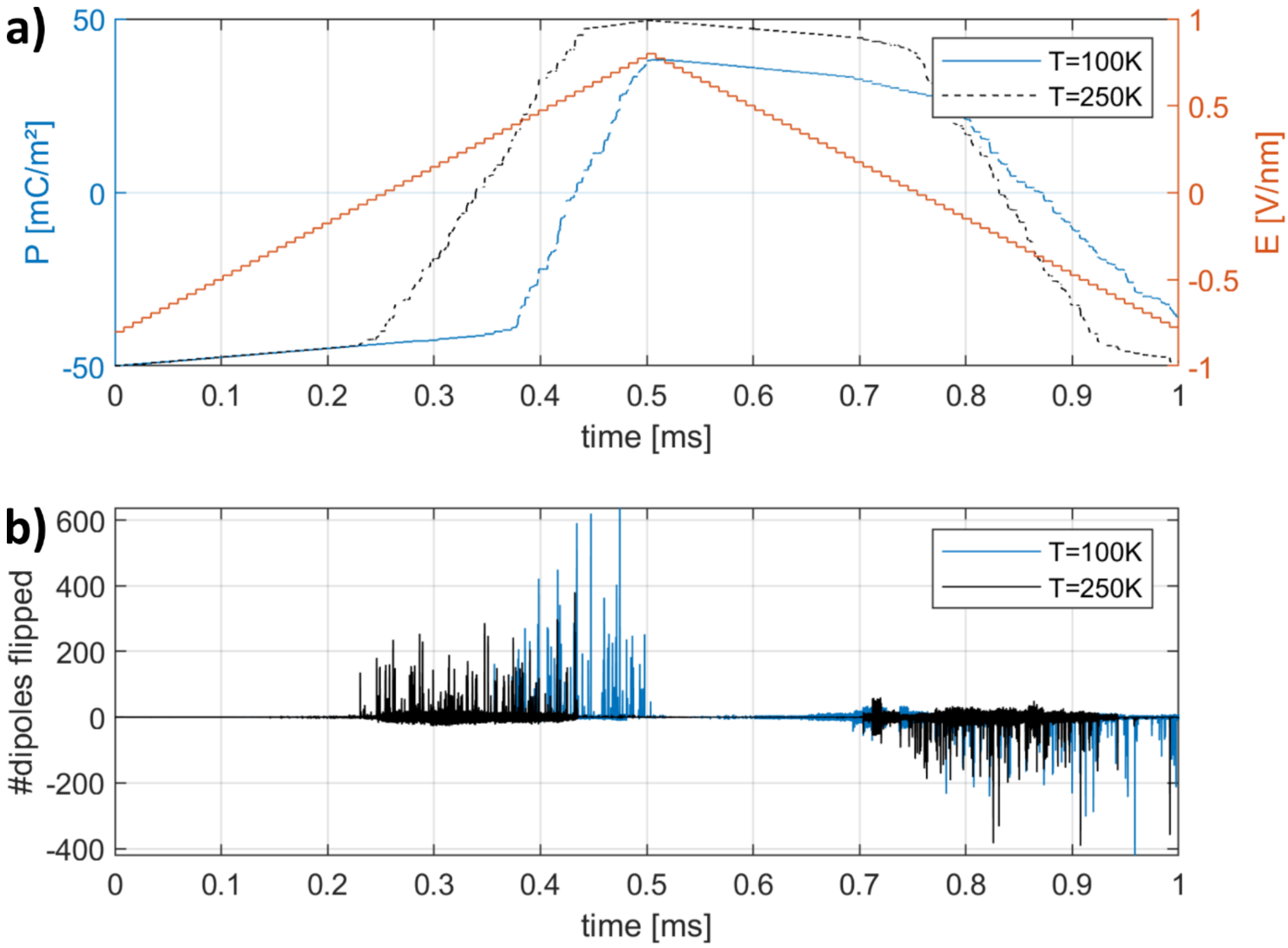

**Figure S4.1:** Comparison of two simulations with different temperatures of 100 and 250 K. **a)** As thermal excitations become less likely, the switching process starts at higher fields and the system is not fully switched at the lowest temperature. **b)** The event sizes for lower temperatures are generally bigger than at high temperatures. The simulation parameters here were $f = 1\,\mathrm{kHz}, n_\mathrm{x} = n_\mathrm{y} = 4, n_\mathrm{z} = 300, N = 10, \sigma_\mathrm{N} = 1$.

## S5. Impact of disorder

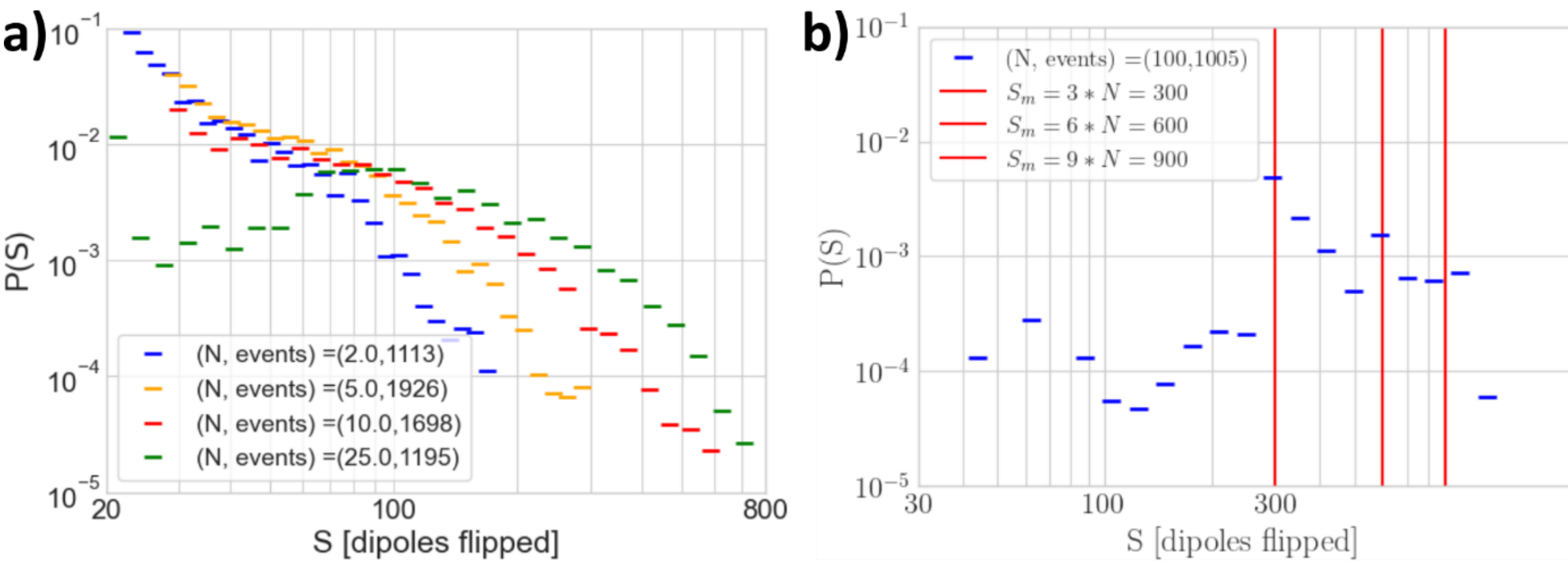

**Figure S5.1:** Probability distributions for different subcolumn sizes within the simulated system. The simulation parameters here were $f = 1$ kHz, $n_x = n_y = 4, n_z = 300, \sigma_N = 1, T = 300$ K. **a)** The increase in the subcolumn size leads to a significant increase of bigger switching events. **b)** Upon reaching too large subcolumn sizes, the power-law behavior vanishes and a clear maximum in the probability appears at multiples of a whole subcolumn being flipped, as indicated by the red vertical lines.

First, the influence of subcolumn sizes on the distribution of event sizes was investigated. The system size was fixed at $n_z = 300$ and the subcolumn size $N$ was varied over a large range while keeping the variance in subcolumn size constant. The resulting distributions are displayed in **Figure S5.1a** and show that with increasing subcolumn sizes, the probability of bigger switching events increases significantly. This can be attributed to the intrinsic character favored by the simulation, as especially for $N = 25$ a clear drop in the probability of events smaller than the flip of one full subcolumn $(S = 3, N = 75)$ is observed. Further increasing the subcolumn size leads to an abrupt process of mainly full subcolumns switching. **Figure S5.1b** shows a clear increase in probabilities of multiples of one full column switching (indicated by the vertical lines) compared to other event sizes for the case $N = 100$, hence giving the distribution a completely different structure. These results are similar to the behavior of the Random-Field-Ising model where the introduced disorder needs to be big enough to prevent the completely intrinsic switching whereas too much disorder in the system leads to small events dominating the switching process.[7] In our case, a mean subcolumn size that significantly exceeds the variance introduces a new characteristic length scale in the system, which is analogous to an increased order. Lastly, the event sizes were also investigated as a function of subcolumn length and, in line with expectation, an increase of the event sizes with the growth of subcolumns was observed, as shown in **Figure S5.2a**. More interesting is that for very small subcolumn sizes ($\leq \sim 5$) the power law exponents start to deviate upwards from the mean field predictions, as depicted in **Figure S5.2b**. This indicates an increasingly large fraction of small events, and corresponds to the Random-Field-Ising limit of too large disorder.

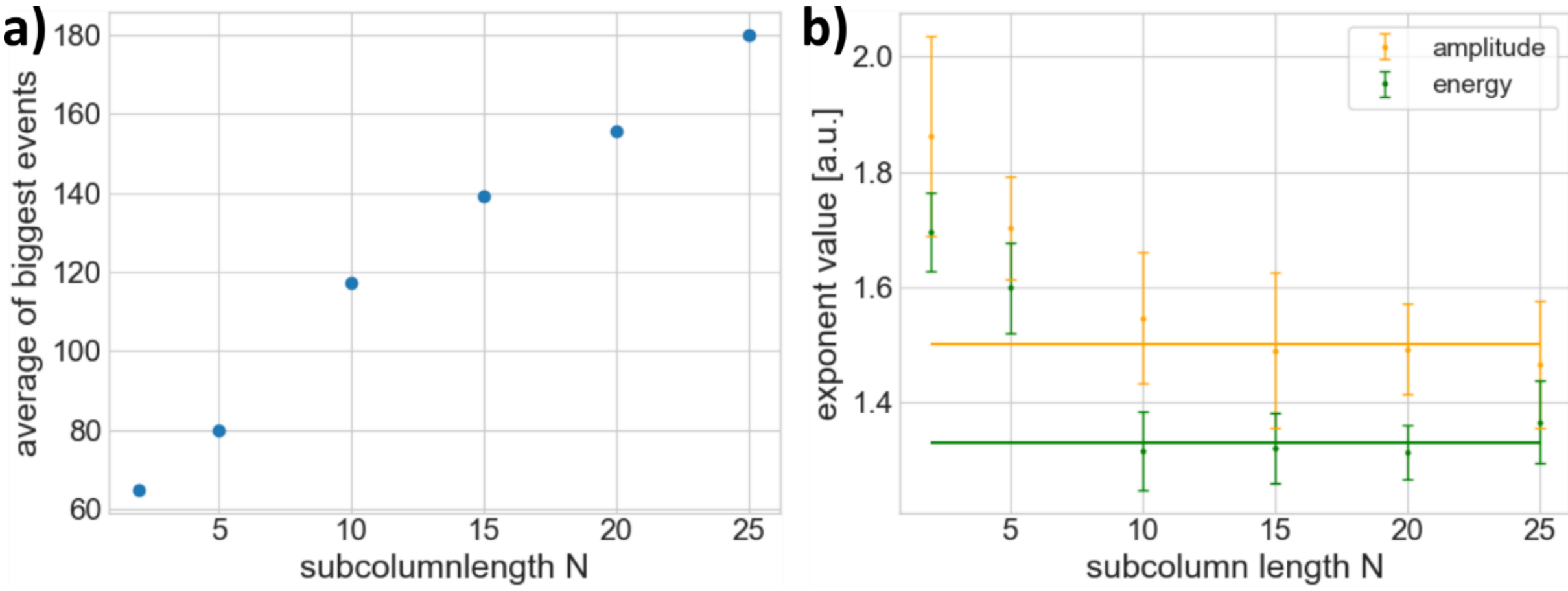

**Figure S5.2:** Effect of disorder on the system. The simulation parameters here were $f = 1\ \text{kHz}, n_x = n_y = 4, n_z = 300, \sigma_N = 1, T = 300\ \text{K}$. **a)** The average of biggest events increases with larger subcolumn length $N$ and thus the defect density decreases for longer subcolumns. **b)** Energy and size exponents are plotted against the subcolumn length $N$ with horizontal lines indicating the mean field predictions. While the exponents increase for smallest subcolumn lengths and hence very high disorder, the exponents are consistent despite the changes of the defect density in the intermediate regime.

For the chosen system with the size $n_z = 300$ and a subcolumn size $N = 10$, the variance of the subcolumn length was varied from $\sigma = 0$ to $\sigma = 5$. A comparison of the distribution and the corresponding extracted exponents is shown in **Figure S5.3** for $T = 130$. Increasing the variance of the subcolumn length leads to a wider distribution of subcolumn sizes. The presence of additional small subcolumns increases the number of events smaller than the average subcolumn length, which does not have a significant effect on the histogram due to overlap with thermal fluctuations and concomitantly the filtering-out via the threshold level. On the other side, the additional bigger columns cause an increase of the probability of bigger events, slightly expanding the range over which the power-law behavior is observed. Importantly, the recovered exponents from the region in which the power-law is present are very close despite the changes in the structure of the distributions. Similar plots to Figure S5.3 were made for temperatures in the range $T = 100$ to $220\ \text{K}$ with increments of 30 K. The results are shown in **Figure S5.4**. No significant effect of the variance of the subcolumn size on the recovered power-law exponents was observed. Thus, the change in variance of the subcolumn length does not seem to have any significant effect on the value and temperature dependence of the power-law exponents, in line with the notion of the system showing self-organized criticality.

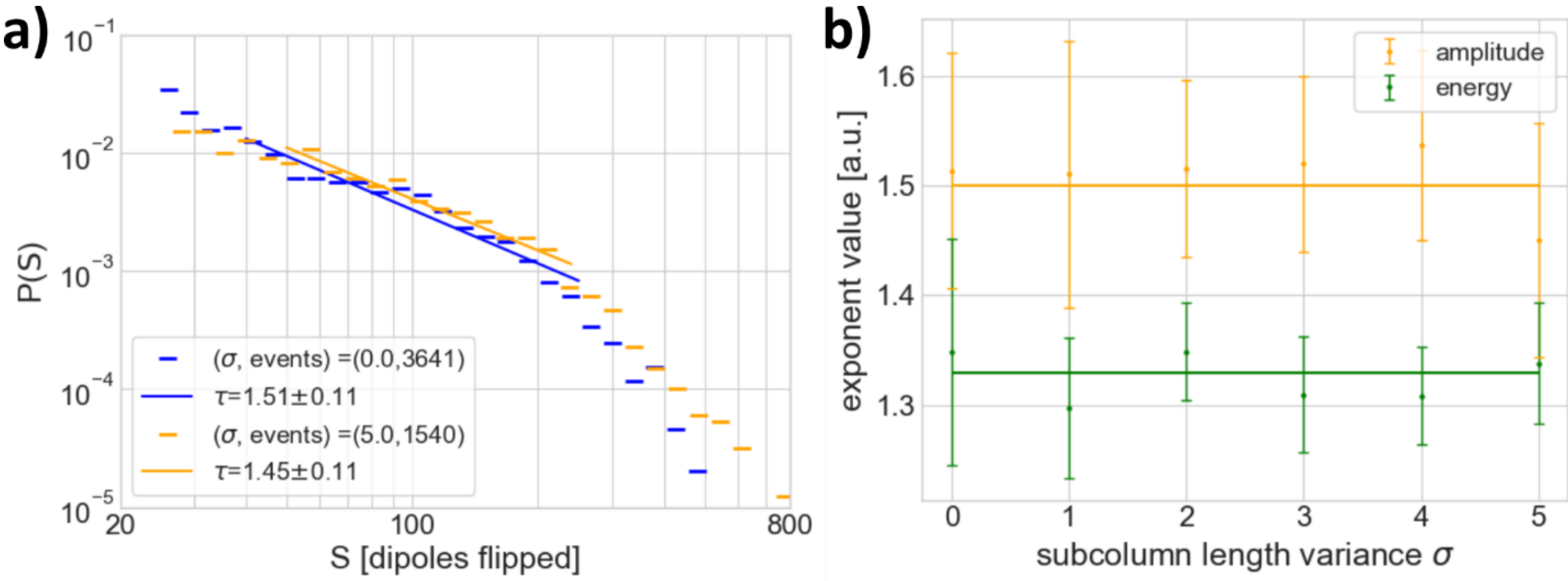

**Figure S5.3:** Effect of the variance of the subcolumn length at a low temperature. The simulation parameters here were $f = 1$ kHz, $n_x = n_y = 4, n_z = 300, N = 10, T = 130$ K. **a)** An increase in the probability of bigger events due to the presence of larger subcolumns is observed. The region in which the power-law behavior is present ($S = 40 - 250$) is not affected significantly. **b)** The extracted power-law exponents stay at the mean field predictions of $\tau = 1.5$ and $\varepsilon = 1.33$ for all investigated variances.

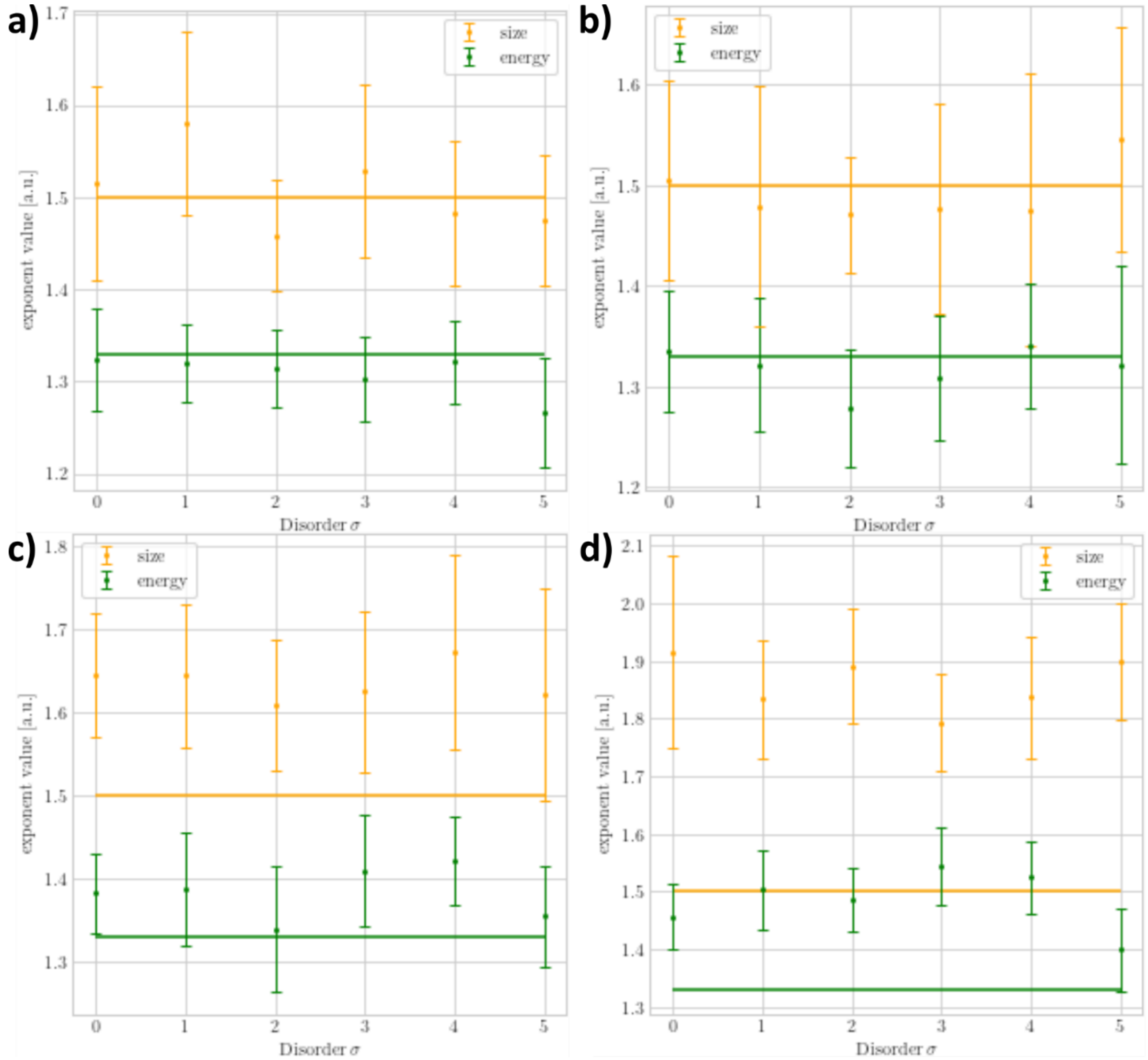

**Figure S5.4:** Extracted power-law exponents as a function of variance of the subcolumn lengths $\sigma_N$ for different lower temperatures as indicated in the graphs. The temperature values were **a)** 100, **b)** 160, **c)** 190 and **d)** 220 K The exponents match the universal critical values (horizontal lines) well for lower temperatures, they exceed them for higher temperatures. In all cases, the exponent values do not show a significant dependency on the applied disorder. The simulation parameters here were $f = 1$ kHz, $n_\mathrm{x} = n_\mathrm{y} = 4, n_\mathrm{z} = 300, N = 10$.

In all simulations shown previously, both chirality and rotation angle of each subcolumn were set to be randomized. To analyze possible effects arising from chirality defects, the corresponding values were separately set to fixed values and the simulations were carried out at $T = 150$ K as the thermal creep showed no significant influence and the whole system could still be switched at this temperature. The chirality was varied such that in one case all defects were set to being chiral, i.e. inverting the handedness with respect to the previous subcolumn, while in the other case the defect would conserve handedness. The resulting distributions for the chirality settings are displayed in **Figure S5.5**. The extracted power-law exponents are presented in **Table S5.1**.

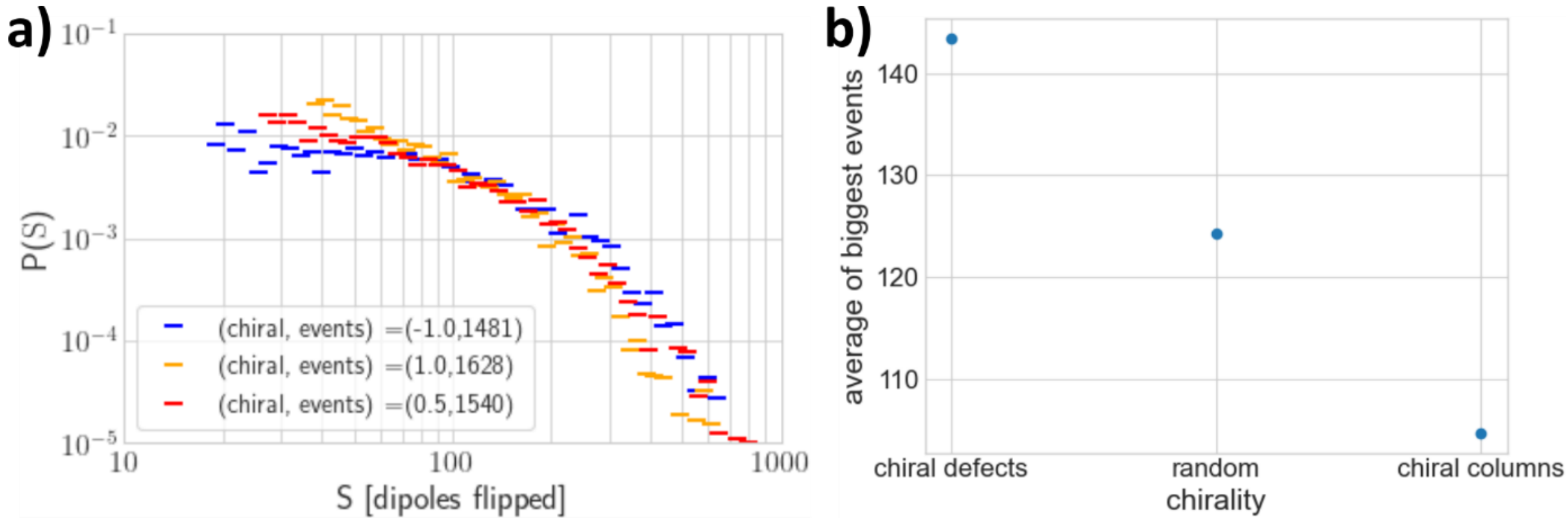

**Figure S5.5:** Effects of different chirality settings on defects between the subcolumns. Both the **a)** probability distributions and **b)** mean event sizes show an increase of the event sizes with the number of chiral defects present in the system. For the chiral settings, -1 corresponds to all defects being chiral, 1 means that no chiral defects are present, and 0.5 means randomized chiral defects. When the number of chiral defects decreases, the mean event size decreases, although the dipoles in fully chiral columns should be better aligned, and hence have stronger interactions. As such, bigger events were expected. A possible explanation is that the energy associated with two oppositely poled subcolumns across a chiral defect is larger than across a non-chiral one. The simulation parameters here were $f = 1$ kHz, $n_x = n_y = 4, n_z = 300, N = 10, \sigma_N = 5, T = 300$ K.

| Parameter | Setting | $\tau$ | $\epsilon$ |
|---|---|---|---|
| Chirality | No chiral defects | 1.71 | 1.43 |
| | Random chirality | 1.54 | 1.34 |
| | Chiral defects | 1.30 | 1.23 |

**Table S2:** Size and energy exponents extracted for different chirality settings from Figure S5.5.

In conclusion, the simulations suggest that a sufficient amount of disorder is needed in order to observe the resemblance of a power-law behavior. However, the type of defects which induce the disorder seem to not matter. Furthermore, adding more defects and hence disorder above the arbitrary threshold to the system seems to not have a significant effect on the exponents until another threshold is reached, where small events dominate.

## S6. Further experimental information

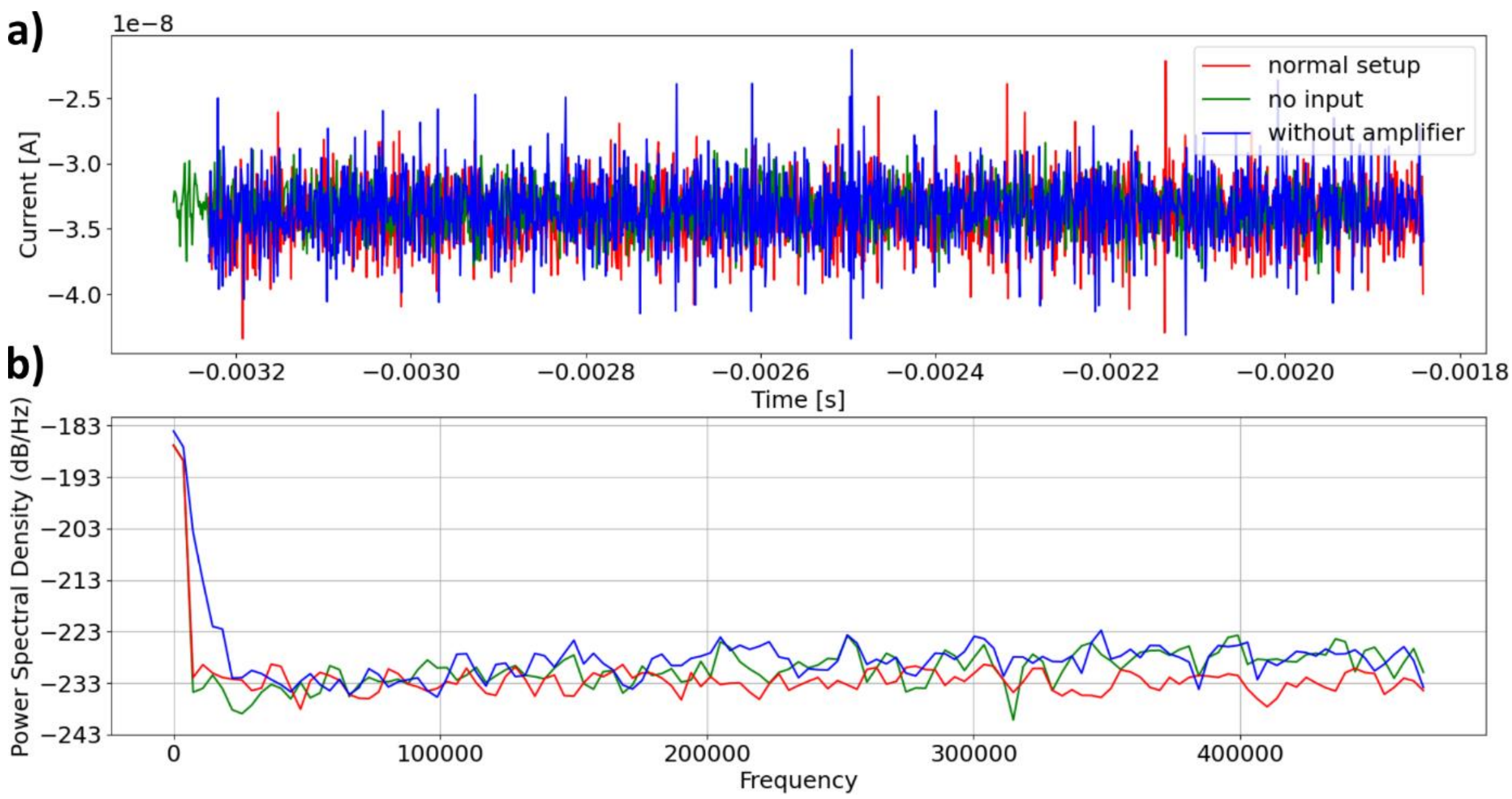

**Figure S6.1:** Noise measurements of the setup with Keysight 33600A Arbitrary Function Generator (AFG) showing **a)** the measured current in the time domain and **b)** the corresponding power spectral density. The noise of the entire setup without a device under test (red line), no input signal (green line, represents the noise floor) and setup without amplifier (blue line) are shown for comparison.

| AFG | Noise number [nA] |
|---|---|
| None (= noise floor of the MFIA/MFLI) | $1.65 \pm 0.03$ |
| Keysight 33600A | $2.57 \pm 0.13$ |
| Tektronix AFG3052C | $4.92 \pm 0.03$ |
| Tektronix AFG1061 | $12.41 \pm 0.14$ |

**Table S6.1:** Obtained noise numbers representing the noise levels of the tested AFGs. The measurements were done by using a reference setup in which the AFG output was connected to a 20× amplifier which led to the sample which was substituted by a dielectric reference capacitor with 10 pF (typical sample capacitance is 6 to 35 pF) and finally into the input of the Zürich Instruments lock-in (MFLI). Thus, the current passing through the capacitor is measured. Since capacitors exhibit a current proportional to the displacement current, only current flow during constant voltage is considered. The MFLI's current range was set to 100 µA with a sampling rate of 937.5 kHz. The noise number is calculated using the standard deviation of the current.

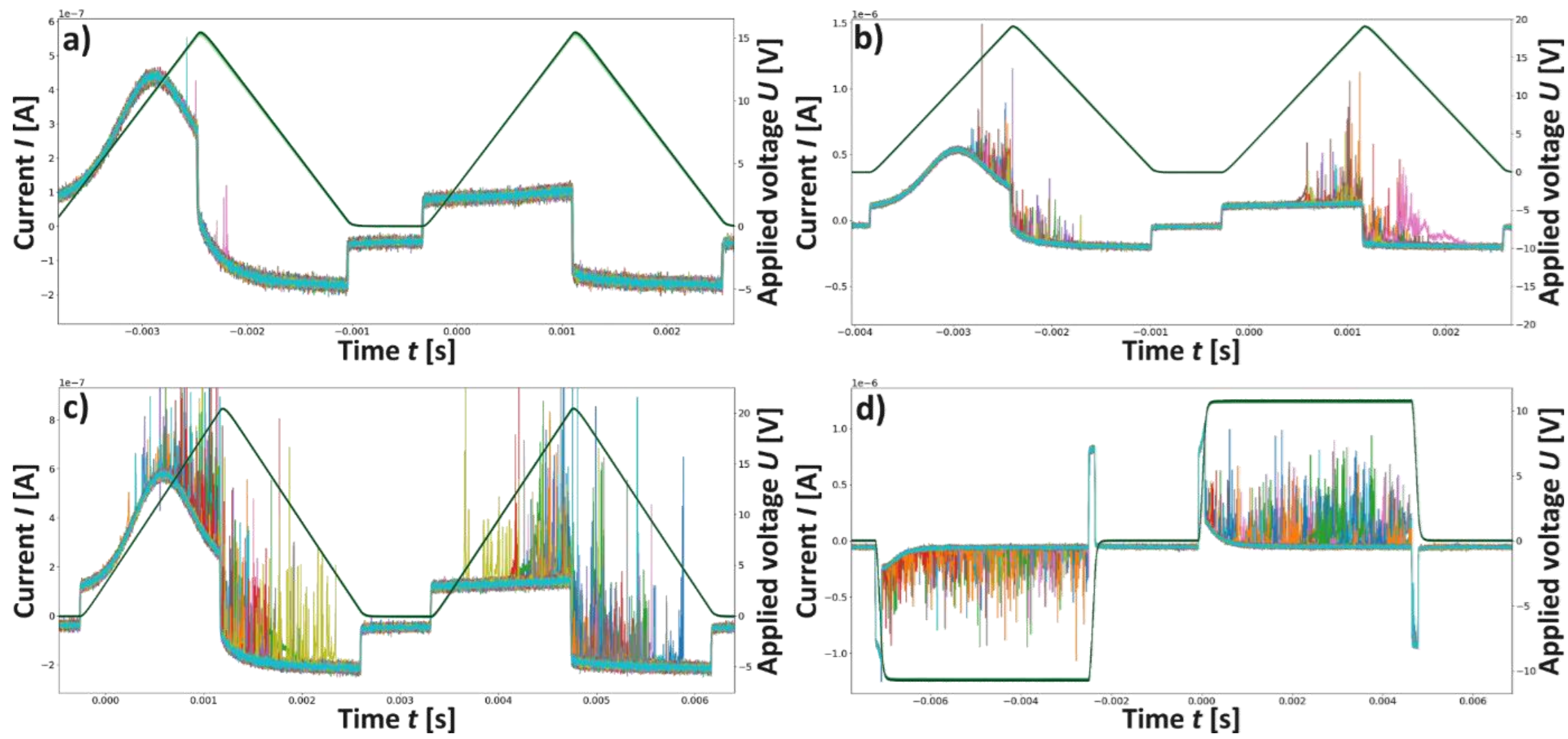

**Figure S6.2:** Applied voltage (dark green line) and the resulting currents (multiple colors) of BTA samples with aluminum and silver electrodes measured at 363 K showing **a)** almost only background noise and **b-d)** artefact spikes for different applied voltages and different waveforms. Here, the applied voltage was either a double triangular signal (a-c) or a double square signal (d) as part of a double wave measurement. The different colors correspond to different measurements and a total of 100 measurements were made in each measurement. The absence of signal (a) is most commonly measured, signals as in (b-d) were occasionally observed. All appear above a threshold voltage of around 12 V and are identical in both up- and down-sweeping signals and for both waves, which is not what should be observed in Barkhausen noise signals for ferroelectric switching (that only happens in the up sweep of the first wave and should be centered around the coercive field). Typically, higher applied voltages led to both a more often and stronger signals, as is indicated by the amount of the peaks. We attribute these peaks to the formation of unstable conductive filaments, which is a general phenomenon in thin film MIM devices.[8,9]

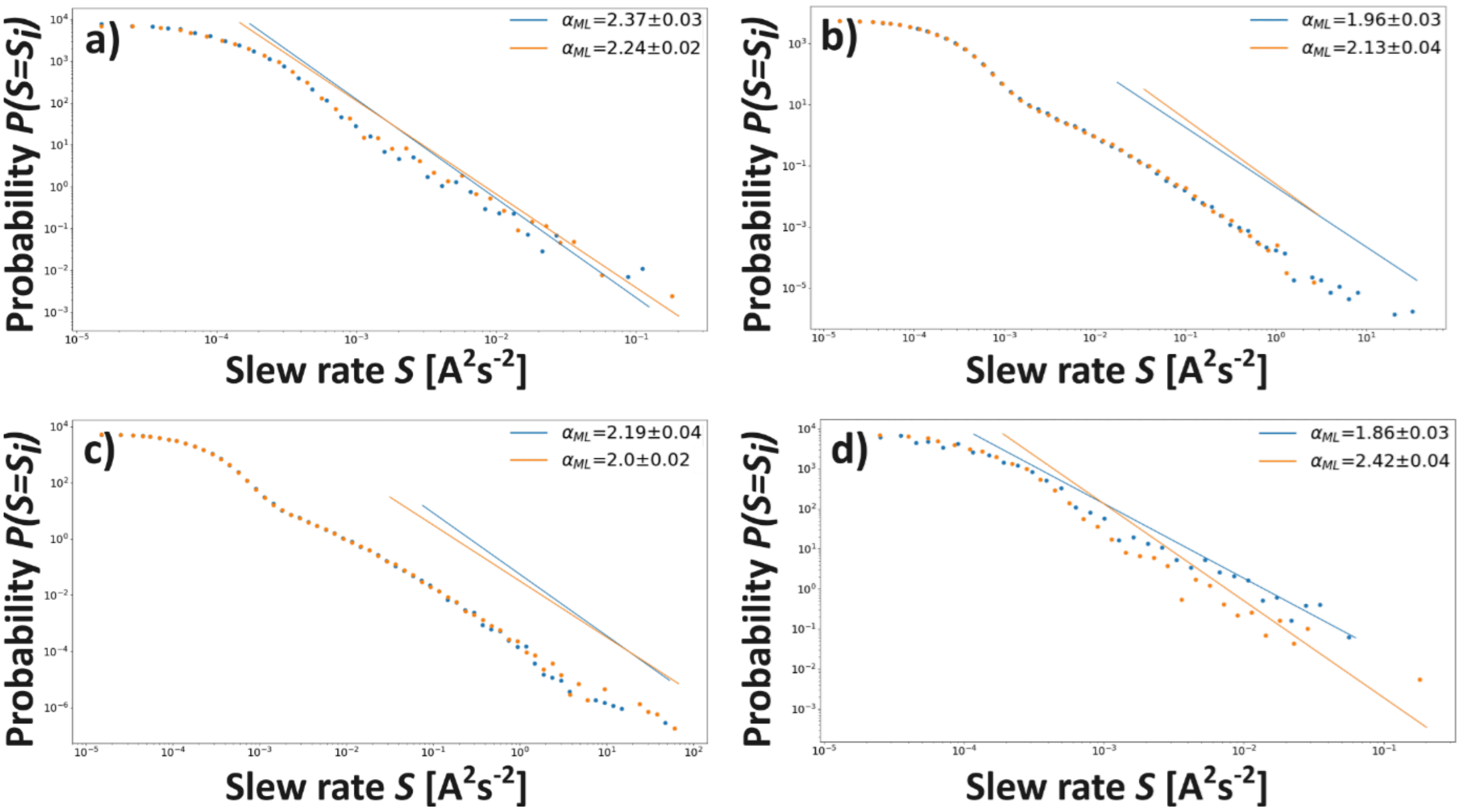

**Figure S6.3:** Probability density distributions and the corresponding exponents of the measurement shown in Figure S6.2. The exponents were fitted by the maximum likelihood method for two different threshold settings, which are indicated in blue and orange. No significant differences between the measurements done for a lower (in blue) and a larger (in orange) threshold setting were observed. In line with the discussion of Figure S6.2, the shape of distribution and the corresponding exponents $(\sim 1.9 - 2.5)$ are not characteristic for Barkhausen noise.

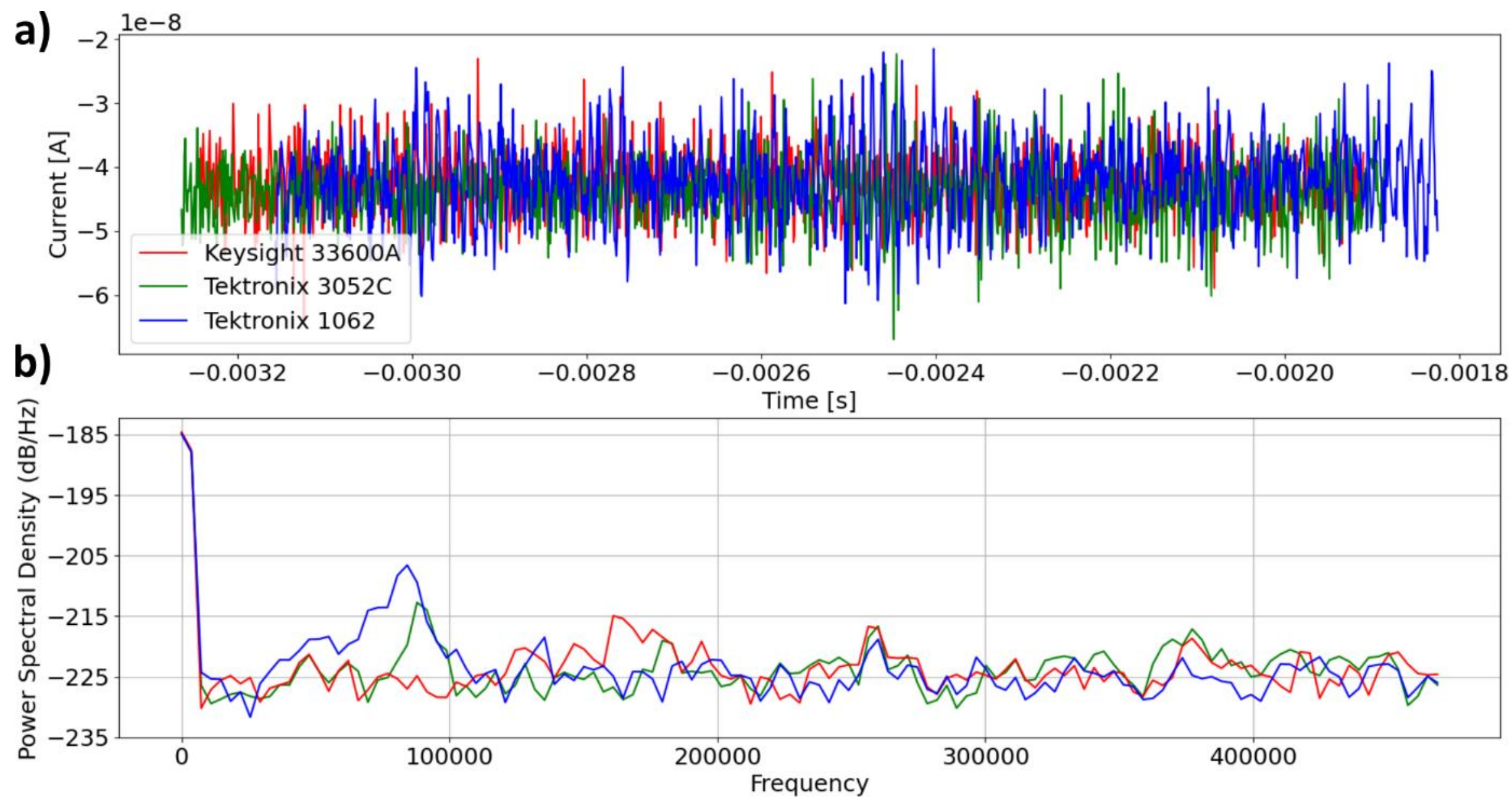

**Figure S6.4:** Noise measurements of the tested AFGs showing **a)** the measured current in the time domain and **b)** the corresponding power spectral density.

## S7. Estimation of the number of switched dipoles

From sample geometry, the total number of switchable dipoles within a material $N_\mathrm{t}$ can be expressed as

$$N_\mathrm{t} = \frac{P_\mathrm{s} A_\mathrm{s} D}{qd}, \qquad (S6.1)$$

where $P_s$ is the saturation polarization, $A_s$ the area and $D$ the thickness of the sample, $q$ the charge and $d$ the size of a microscopic dipole. Thus, the number of dipoles in the material that are switched by a change in charge $N_s$ are given by

$$N_\mathrm{s} = \frac{QD}{qd}, \qquad (S6.2)$$

with the total charge $Q$, so that $QD$ is the macroscopic dipole moment. The total charge can be obtained from the switching current $I(t)$ via $Q = \int I(t)dt$.

Taking a typical out-of-plane device with an active area of 0.25 mm x 0.25 mm and a film thickness of 400 nm, assuming 0.34 nm as π-stacking and 1.94 nm as intermolecular distances [10], results in a maximum column height of around 1180 BTA-$C_{10}$ molecules between the top and bottom electrodes. Further assuming a simple hexagonal 2D Bravais lattice, the total number of dipoles in the BTA sample can be calculated via

$$N_\mathrm{t}^\mathrm{g} = n_\mathrm{d} N_\mathrm{ch} c \frac{A_\mathrm{s}}{A_\mathrm{uc}} = n_\mathrm{d} N_\mathrm{ch} N_\mathrm{c} \qquad (S6.3)$$

with $n_\mathrm{d} = 3$ as the number of dipoles per molecule, $N_\mathrm{ch} \approx 1180$ as the number of molecules in a column, $c = 4$ as the coordination number, $A_\mathrm{s}$ as the area and $A_\mathrm{uc} = \frac{3\sqrt{3}}{2} a^2$ the area of the hexagonal unit cell with $a = 1.94$ nm as the intermolecular distance. This results in $\sim 3 \cdot 10^{13}$ BTA molecules and hence a total of $\sim 9 \cdot 10^{13}$ dipoles.

Alternatively, using that BTA-$C_{10}$ has a remnant polarization of around 45 mC $m^{-2}$ at the measured temperature [11], **Equation S6.1** yields a total of $\sim 8.4 \cdot 10^{13}$ dipoles which is in good agreement with the previous estimation.

In the kMC simulation, it was concluded that assuming a constant number of dipoles the average size of dipole avalanches flipping mainly depends on the column length which indicates that the columns are only weakly coupled. It was established that for 700 BTA molecules in a column and a width of four columns in both x- and y-directions, 500 dipoles per event were the maximal event size and the power-law onset occurred between 100 and 200 dipoles (see Figure S3.1). Assuming an event size of 200 dipoles and that the experimentally measured sample consists of multiple such simulated regions, a rough estimate indicates that the measured BTA sample consists of $\sim 1.6 \cdot 10^9$ regions.

In order to achieve a measurable current (in our case $\sim 4$ nA for 1 µs time resolution; 938 kHz sampling rate), a total of $\sim 6 \cdot 10^5$ regions have to simultaneously provide a switching event. Due to the weak coupling between the columns and avalanches only occurring within a column, this is very unlikely to happen, implying that the actually occurring switching events will be far below the resolution threshold of the used setup.

## Supplementary references